\documentclass[twocolumn]{aastex631}

\usepackage{amsmath} 
\usepackage{soul} 
\usepackage{multirow} 
\usepackage{bm} 
\usepackage{amsmath} 
\usepackage{xspace} 
\usepackage{gensymb} 
\usepackage{float}  

\usepackage{newunicodechar,graphicx}
\usepackage{multirow, bigdelim, makecell, booktabs}
\PassOptionsToPackage{hyphens}{url}\usepackage{hyperref}
\usepackage{mhchem} 

\newcommand{\eureka}[1]{\texttt{Eureka!}{#1}}
\newcommand{\tiberius}[1]{\texttt{Tiberius}{#1}}
\newcommand{\exotic}[1]{\texttt{ExoTIC-JEDI}{#1}}

\begin{document}

\title{A dark, bare rock for TOI-1685~b from a JWST NIRSpec G395H phase curve}

\correspondingauthor{Rafael Luque}
\email{rluque@uchicago.edu}

\author[0000-0002-4671-2957]{Rafael Luque}
\altaffiliation{NHFP Sagan Fellow}
\affiliation{Department of Astronomy and Astrophysics, University of Chicago, Chicago, IL 60637, USA}

\author[0000-0002-0508-857X]{Brandon Park Coy}
\affiliation{Department of the Geophysical Sciences, University of Chicago, Chicago, IL 60637, USA}

\author[0000-0002-6215-5425]{Qiao Xue}
\affiliation{Department of Astronomy and Astrophysics, University of Chicago, Chicago, IL 60637, USA}

\author[0000-0002-9464-8101]{Adina~D.~Feinstein}
\altaffiliation{NHFP Sagan Fellow}
\affiliation{Department of Physics and Astronomy, Michigan State University, East Lansing, MI 48824, USA}

\author[0000-0003-0973-8426]{Eva-Maria Ahrer}
\affiliation{Max-Planck-Institut f\"{u}r Astronomie, K\"{o}nigstuhl 17, 69117 Heidelberg, Germany}

\author[0000-0001-6516-4493]{Quentin Changeat}
\affiliation{Kapteyn Institute, University of Groningen, 9747 AD Groningen, NL}
\affiliation{Department of Physics and Astronomy, University College London, WC1E 6BT London, UK}

\author[0000-0002-0659-1783]{Michael Zhang}
\affiliation{Department of Astronomy and Astrophysics, University of Chicago, Chicago, IL 60637, USA}

\author[0000-0002-6721-3284]{Sarah E. Moran}
\altaffiliation{NHFP Sagan Fellow}
\affiliation{NASA Goddard Space Flight Center, Greenbelt MD 20771, USA}

\author[0000-0003-4733-6532]{Jacob L.\ Bean}
\affiliation{Department of Astronomy and Astrophysics, University of Chicago, Chicago, IL 60637, USA}

\author[0000-0002-1426-1186]{Edwin Kite}
\affiliation{Department of the Geophysical Sciences, University of Chicago, Chicago, IL 60637, USA}

\author[0000-0003-4241-7413]{Megan Weiner Mansfield}
\affiliation{School of Earth and Space Exploration, Arizona State University, Tempe, AZ 85281, USA}
\affiliation{Department of Astronomy, University of Maryland, College Park, MD 20742, USA}

\author[0000-0003-0987-1593]{Enric Pall\'e}
\affiliation{Instituto de Astrof\'isica de Canarias (IAC), 38200 La Laguna, Tenerife, Spain}
\affiliation{Deptartamento de Astrof\'isica, Universidad de La Laguna (ULL), 38206 La Laguna, Tenerife, Spain}

\begin{abstract}

\noindent We report JWST NIRSpec/G395H observations of TOI-1685~b, a hot rocky super-Earth orbiting an M2.5V~star, during a full orbit. We obtain transmission and emission spectra of the planet and characterize the properties of the phase curve, including its amplitude and offset. The transmission spectrum rules out clear H$_2$-dominated atmospheres, while secondary atmospheres (made of water, methane, or carbon dioxide) cannot be statistically distinguished from a flat line. The emission spectrum is featureless and consistent with a blackbody-like brightness temperature, helping rule out thick atmospheres with high mean molecular weight. Collecting all evidence, the properties of TOI-1685~b are consistent with a blackbody with no heat redistribution and a low albedo, with a dayside brightness temperature $0.98\pm0.07$ times that of a perfect blackbody in the NIRSpec NRS2 wavelength range (3.823--5.172\,µm). Our results add to the growing number of seemingly airless M-star rocky planets, thus constraining the location of the `Cosmic Shoreline'. 

Three independent data reductions have been carried out, all showing a high-amplitude correlated noise component in the white and spectroscopic light curves. The correlated noise properties are different between the NRS1 and NRS2 detectors -- importantly the timescales of the strongest components (4.5 hours and 2.5 hours, respectively) -- suggesting the noise is from instrumental rather than astrophysical origins. We encourage the community to look into the systematics of NIRSpec for long time-series observations.

\end{abstract}



\section{Introduction} \label{sec:intro}

Whether rocky planets around M~dwarfs can retain significant atmospheres remains unclear.  Emission observations thus far have all been consistent with the `bare rock' scenario \citep{kreidberg19,Crossfield22,greene23,zieba23,zhang24,xue24,mansfield24,Wachiraphan24}, and transmission observations have failed to make confident detections of any molecular features \citep{Lustig-Yaeger2023NatAs...7.1317L,Alderson2024,Wallack2024}, with stellar contamination potentially a source of the few non-flat spectra seen to date \citep{may2023,MoranStevenson2023,lim2023}.  The `Cosmic Shoreline' hypothesis \citep{zahnle2017} predicts that most rocky planets around M~dwarfs cannot retain significant atmospheres due to enhanced XUV radiation from their low-mass host stars.  However, late-stage outgassing or cometary impacts have been proposed as pathways to form secondary atmospheres for M-star rocky planets (e.g., \citealt{kral18,kite20}).

Spectroscopic phase curves are a powerful tool for characterizing tidally-locked hot rocky exoplanets.  Unlike eclipse observations, phase curves can directly constrain the level of heat redistribution to the nightside, constraining the thickness of an atmosphere if one exists (e.g., \citealt{kreidberg19,zhang24}). 
In the case of an airless planet, spectroscopic emission observations can in principle be used to probe the planet's surface composition (e.g., \citealt{hu12,whittaker22}).  Spectroscopic phase curves also provide important phase-resolved information about surface geology, possibly constraining the planet's surface roughness (amount of cratering) and/or level of space weathering \citep{lyu24,tenthoff24}. 

Here, we present JWST NIRSpec/G395H observations of the M-dwarf planet TOI-1685~b during a full orbit. Originally proposed as a bona fide water world candidate due to its high equilibrium temperature ($T_{\rm eq} \sim 1000\,\mathrm{K}$) and low bulk density \citep{bluhm2021,luquepalle22}, a recent study by \citet{Burt2024arXiv240514895B} confirmed the planet to be a hot super-Earth with an Earth-like density orbiting a relatively inactive M2.5V star. The target has been selected by two other JWST Cycle 2 programs \citep{go4098,go4195} focusing on characterizing its potential atmosphere in transmission. This program covers the full orbit of the planet, enabling the detection of the planet's thermal emission directly and constraining the presence or absence of an atmosphere on the planet more tightly than transit observations alone. 

The paper is structured as follows. A description of the observations and independent data reductions is shown in Sect.~\ref{sec:data}. The modeling and analysis of the white and spectroscopic light curves are described in Sect.~\ref{sec:analysis}, including the characterization of the correlated noise component found in the phase curve residuals. Section~\ref{sec:discussion} contains the analyses of the transmission and emission spectra of the planet and the properties of the phase curve, which jointly suggests that TOI-1685~b lacks an atmosphere and is consistent with a dark, bare rock.


\section{Observations and data reduction} \label{sec:data}

We observed TOI-1685~b with the NIRSpec instrument on JWST \citep[][]{Birkmann2022A&A...661A..83B, Jakobsen2022A&A...661A..80J} on 14-15 February 2024 between 23:37 and 18:36 UTC as part of program GO-3263 \citep[PI: Luque;][]{2023jwst.prop.3263L}. We used the Bright Object Time Series (BOTS) mode with the G395H grating. The observations were designed similarly to most phase curves: starting 1\,hour before the ingress of a secondary eclipse up until the end of the following eclipse. The uninterrupted observation lasted 18.97\,hours and consisted of 4450 integrations using 16 groups per integration. 

To ensure the robustness and accuracy of our results, we perform multiple independent data reductions using three different well-tested and widely-used open-source pipelines, namely \eureka\ \citep[][]{eureka}, \tiberius\ \citep{Kirk2017,Kirk2021}, and \exotic\ \citep{alderson22, Alderson2023}. An overview of the high-level steps of each data reduction is given in the following subsections.

\subsection{\texttt{Eureka!}} \label{subsec:eureka}

\eureka\ \citep{eureka} is an end-to-end pipeline for analyzing HST and JWST transiting exoplanet data divided into multiple, independent stages. The first two Stages act as a wrapper of the \texttt{jwst} pipeline \citep[v.1.14.0 in this analysis;][]{2024zndo..10870758B}. We followed the default settings in Stage 1 except for the group-level background subtraction, which we applied to account for 1/$f$ noise \citep[e.g.,][]{Lustig-Yaeger2023NatAs...7.1317L}, and increasing the jump step detection threshold from 4 (default) to 10$\sigma$. In Stage 2, we skip the flat fielding and photometric calibration steps as the conversion to physical flux units is unnecessary for the relative flux measurements we are interested in and only introduces noise to the extracted light curves.

Stage 3 performs background subtraction and optimal spectral extraction, generating a time series of 1D spectra. First, \eureka\ corrects the curvature of NIRSpec G395H spectra by shifting the detector columns by integer pixels, bringing the peak of the distribution of the counts in each column to the center of the subarray. Then, we applied a column-by-column background subtraction to the region more than 7 pixels from the trace center with a 7$\sigma$ outlier threshold in both the spatial and temporal directions. After, we construct a median frame for optimal spectral extraction \citep{Horne1986PASP...98..609H} and use an aperture half-width size of 3 pixels on either side of the center pixel with a 15$\sigma$ threshold for outlier rejection in both detectors NRS1 and NRS2. We tested multiple combinations of aperture and background apertures and this one (3-pix half-width for science spectra and 7-pix half-width for background) minimizes the median absolute deviation of the resulting white light curves. In Stage 4, we generate white (summing the flux over 2.844-3.715\,µm for NRS1 and 3.823-5.712\,µm for NRS2) and spectroscopic light curves at two different resolutions: a very low-resolution spectrum to study the planet's thermal emission (5 spectroscopic channels per detector; 0.17 and 0.27\,µm wide for NRS1 and NRS2, respectively, $R \sim 15-21$) and a higher resolution spectrum to study the planet's atmosphere in transmission (50-pixel binned, 0.033\,µm wide, $R \sim 87-152$). 

\subsection{\texttt{Tiberius}} \label{subsec:tiberius}

For providing an additional independent reduction of the data we used open-source pipeline \texttt{Tiberius} which has been developed for ground-based data reduction \citep{Kirk2017,Kirk2021}, but adapted and used widely for JWST data analyses \citep[e.g.][]{JWSTERS2023,Rustamkulov2023,Alderson2023,Benneke2024TOI270,Kirk2024}. We start our analysis with the uncalibrated FITS files and run the default \texttt{jwst} pipeline steps for NIRSpec time series observations with the exception of the \texttt{photom\_step} and \texttt{jump\_step}. Before running the ramp-fitting we use the \texttt{Tiberius} group-level background subtraction, a column-by-column median subtraction to reduce 1/$f$ noise. We also chose to run the \texttt{assign\_wcs\_step} and \texttt{extract\_2d\_step} steps from the \texttt{jwst} pipeline to extract the wavelength solutions for our integrations. Before extracting the time series spectra, we performed a cosmic pixel outlier cleaning step, where we replaced pixels with values larger than $5\sigma$ compared to the running median across 3 frames for each pixel. 

The spectral extraction was then carried out using a full width of 10\,pixels after performing a column-by-column background subtraction using the remaining area of the frame located $>4$\,pixels from the edge of the aperture and a linear polynomial. The center of the trace was identified at each column using a Gaussian. We extracted spectra from 700 -- 2400 pixels for NRS1 and 8 -- 2040 pixels for NRS2, which were then fed into \texttt{Eureka!}'s Stage\,4 for further analysis of the phase curve. 

\subsection{\texttt{ExoTIC-JEDI}} \label{subsec:exotic}

We used Exoplanet Timeseries Characterisation - JWST Extraction and Diagnostic Investigator \citep[\exotic;][]{alderson22, Alderson2023}\footnote{\url{https://github.com/Exo-TiC/ExoTiC-JEDI}} for a third independent reduction of the same data set. \exotic\ is an open-source pipeline that performs end-to-end extraction and analysis of JWST time-series observations from \texttt{uncal} files through fitting spectroscopic light curves. The NRS1 and NRS2 data are fit independently throughout this pipeline. We use the same setup, as described below, for each detector.

We start our analysis with the uncalibrated data and run a modified version of Stage 1 of the \texttt{jwst} pipeline. We perform linearity, dark current, and saturation corrections, and use a ramp jump rejection threshold of 15. We perform a custom bias subtraction. The custom bias frame is created by computing the median pixel value of every integration in the first group. This new bias is subtracted from each group. Additionally, \exotic\ performs a custom 1/$f$ noise correction on the group level, which is performed during Stage 1. This additional correction masks the spectral trace and subtracts the median pixel value of non-masked pixels from each column. We ran the \exotic\ reduction including and excluding this custom noise removal. The resulting transmission spectra were consistent within $1\sigma$ for this data set. We present the results of the \exotic\ reduction which does not include this custom noise removal step. Our Stage 2 analysis follows the standard process of the \textit{jwst} pipeline; the 2D wavelength map is created at this time and is used to obtain our wavelength solution.

We extract the 1D stellar spectra during Stage 3. Before extraction, we perform an additional outlier removal step, adopting an outlier threshold of $4\sigma$. While $4\sigma$ may seem restrictive, we find the resulting stellar spectra to be consistent with the results of the \eureka\ pipeline. Additionally, any pixels flagged in the \texttt{jwst} quality flag images are interpolated by using the median value of the neighboring four pixels across the same row (two pixels on each side of the bad pixel). We perform this procedure for bad pixels that are flagged as do not use, dead, hot, saturated, low quantum efficiency, or no gain. We perform outlier rejection for additional spurious pixels which may have appeared due to i.e., cosmic ray hits. We correct for 1/$f$ noise on the integration level. The 1/$f$ noise is corrected by masking the spectral trace and calculating the median pixel value of the column. This value is then subtracted from the column and performed on each integration. Finally, the spectral trace is identified by fitting a Gaussian to each column. The mean of the Gaussian is identified as the center of the trace and fitted with a fourth-order polynomial. Since the spectral trace does not cover the entire image for NRS1, we used columns 700 -- 2042 for this analysis. Using this new trace, we performed a simple box extraction with a width of 12-pixels to extract the 1D stellar spectra.


\section{Modelling} \label{sec:analysis}

\subsection{White light phase curve} \label{subsec:white}

\begin{figure*}[t]
    \centering
    \includegraphics[width=0.99\linewidth]{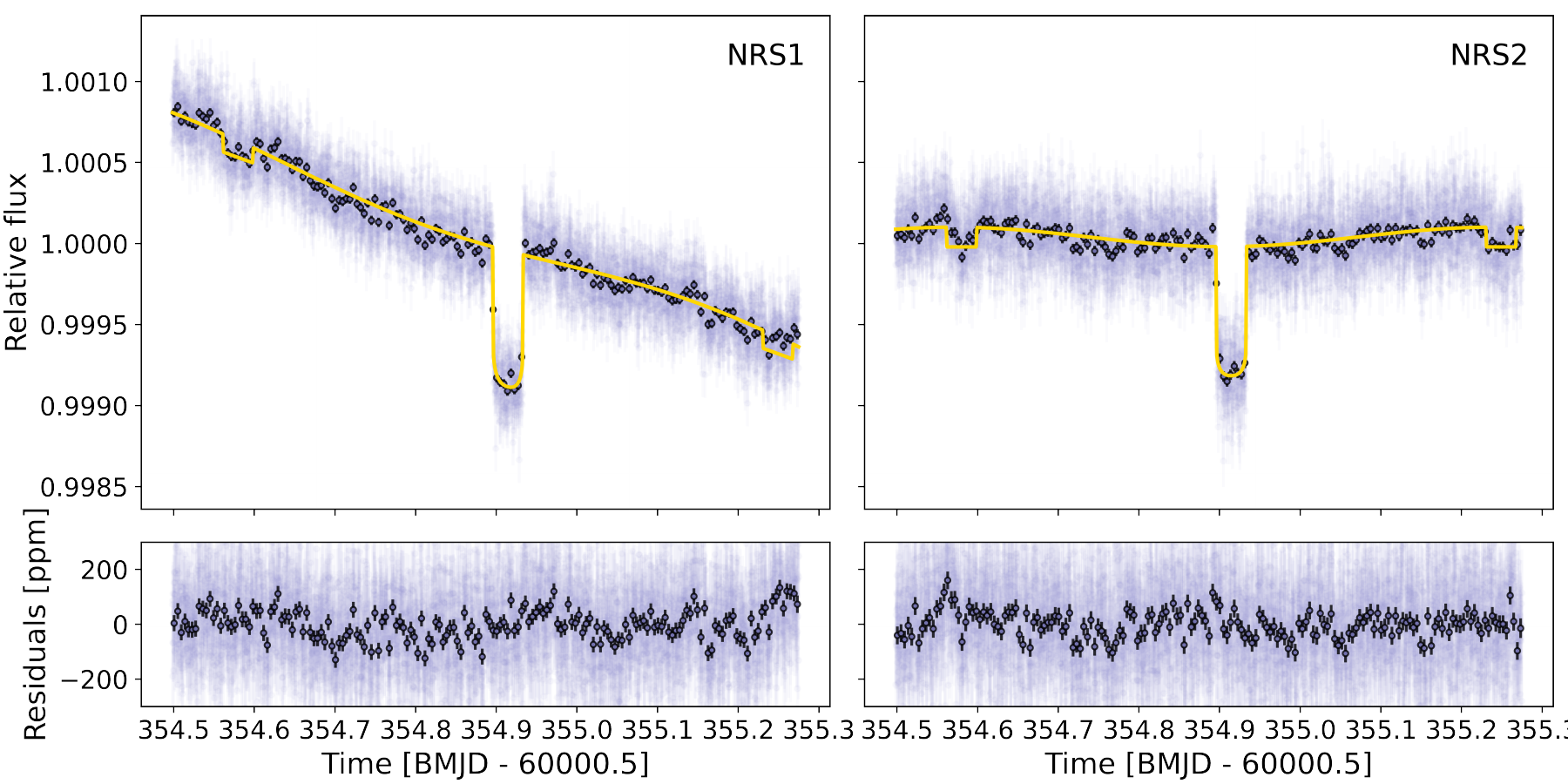}
    \caption{The white-light phase curve of TOI-1685~b from 2.844-3.715\,µm (NRS1, left column) and 3.823-5.172\,µm (NRS2, right column) from JWST. Top: data from the \eureka\ reduction (purple) and planet flux model fit to the data (gold). Bottom: residuals of the fit. Circles show binned data every 25 integrations to improve visualization. }
    \label{fig:phasecurve}
\end{figure*}

\begin{figure*}[t!]
    \centering
    \includegraphics[width=0.49\linewidth]{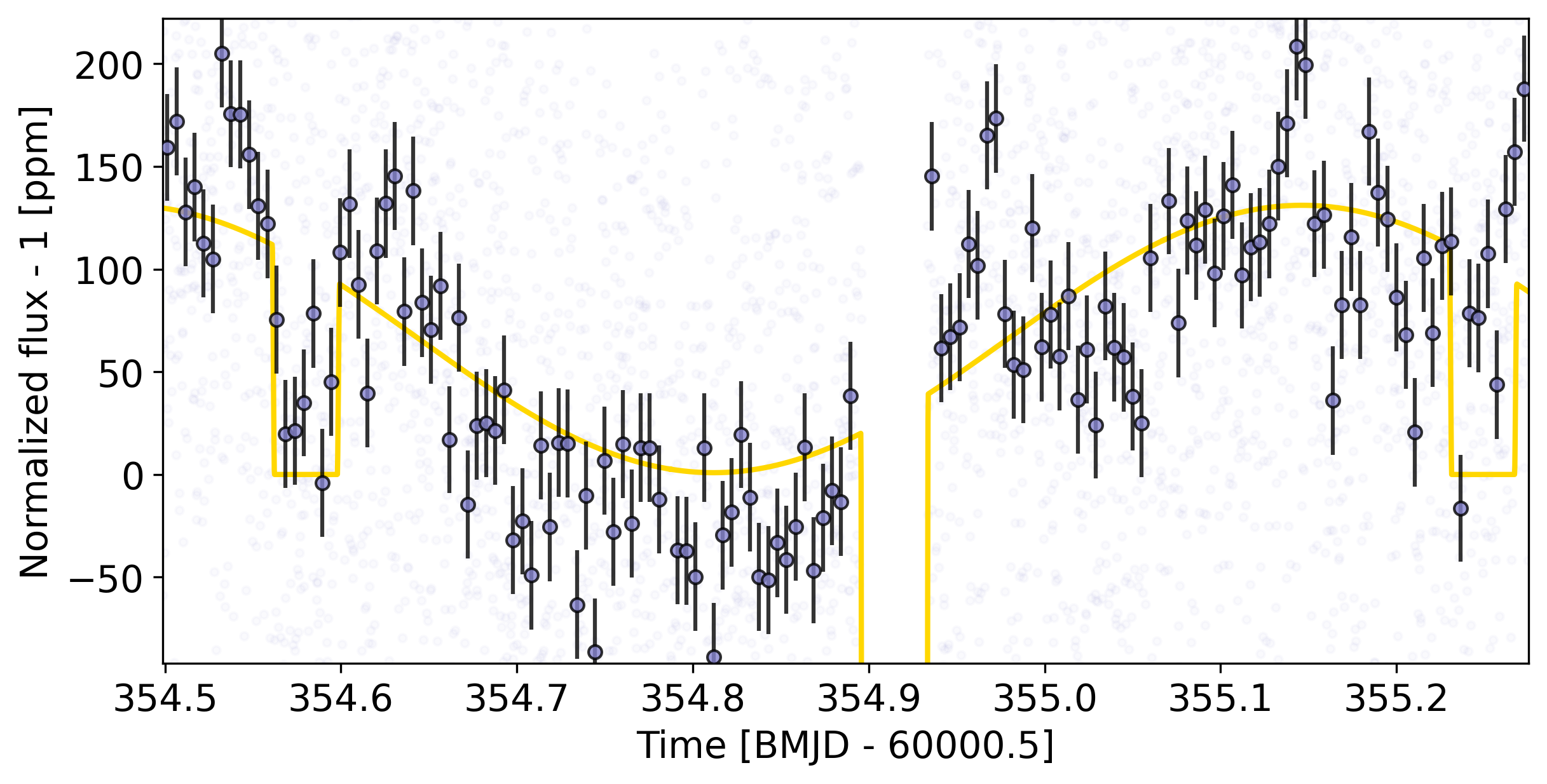}
    \includegraphics[width=0.49\linewidth]{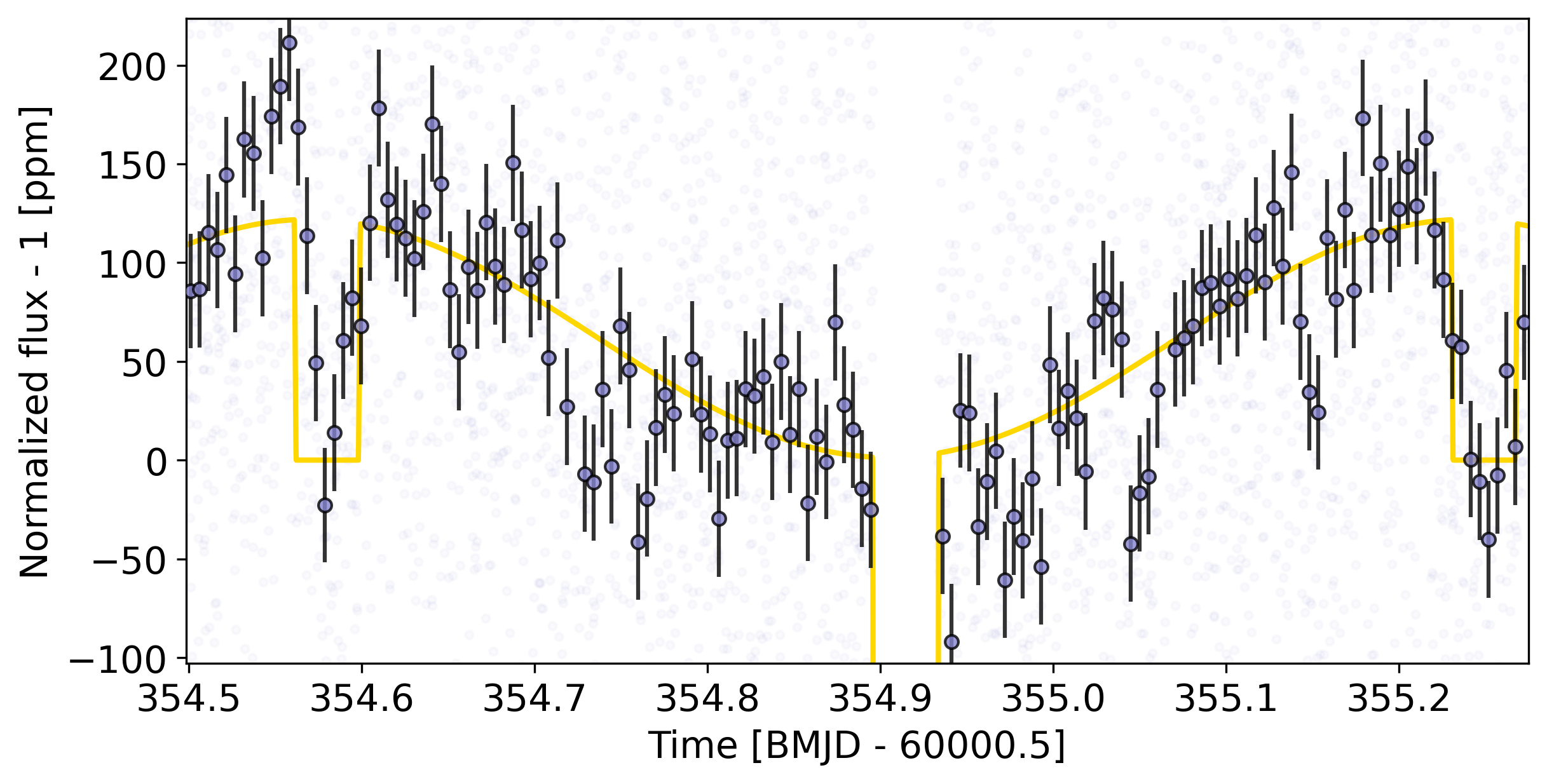}
    \caption{As in Fig.~\ref{fig:phasecurve}, but correcting for the instrument baseline trend of NRS1. Left is NRS1 and right is NRS2. The y-axis is zoomed in to see the planetary emission signal. }
    \label{fig:phasecurve_zoom}
\end{figure*}

\begin{figure}[t]
    \centering
    \includegraphics[width=1.0\linewidth]{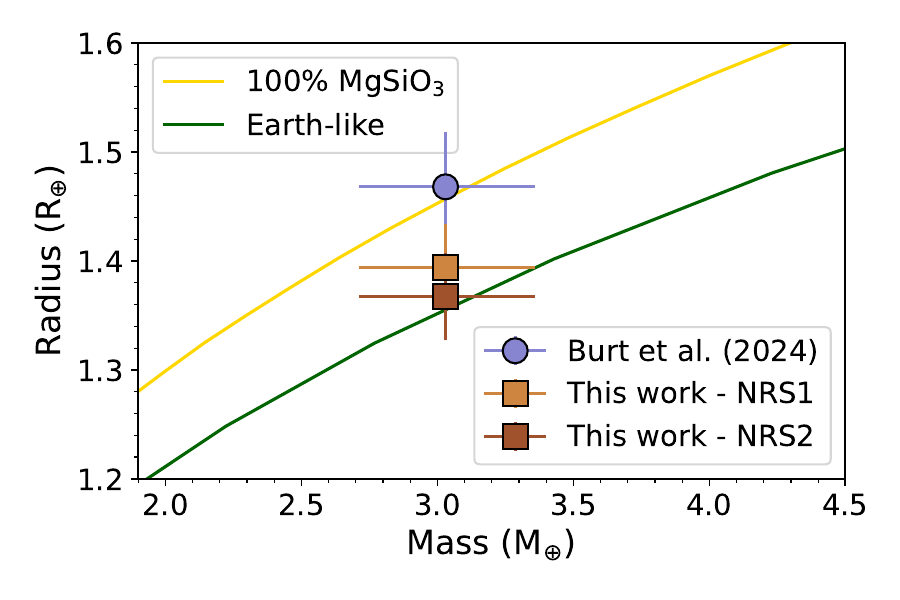}
    \caption{Position of TOI-1685~b in a mass-radius diagram. Internal composition models of rocky planets from \citet{Zeng2016ApJ...819..127Z}. The change in radius from the highest precision of JWST observations compared to TESS makes the density of TOI-1685~b even more consistent with Earth than previously reported.}
    \label{fig:mass_radius}
\end{figure}

As described above, we generated white light curves for each detector NRS1 and NRS2 by summing the flux over 2.844-3.715\,µm and 3.823-5.172\,µm, respectively. The resulting light curves from the \eureka\ reduction are shown in Fig.~\ref{fig:phasecurve}. The NRS1 data shows a steady downward linear trend of 76\,ppm per hour and short-period variability with a timescale of approximately 4\,hours and peak-to-peak amplitude of 200\,ppm. We do not find evidence for this linear trend in NRS2, but we find short-period variability with a timescale of approximately 2\,hours and peak-to-peak amplitude of 200\,ppm. We do not find any evidence for an exponential ramp at the beginning of the observations. All three data reductions provide very consistent results as discussed below in Sect.~\ref{subsec:rednoise}.

We modeled these light curves independently within Stage 5 of \eureka using the \texttt{batman} package \citep{batman}. The full phase curve model assumes a circular orbit\footnote{Models with free eccentricity find $e=0.0011^{+0.0013}_{-0.0007}$ and $\omega = 60^{+31}_{-40}$\,deg, consistent with a circular orbit given the Lucy-Sweeney bias on the eccentricity, which is inherently constrained to be a positive value \citep{Lucy1971}.} and includes a transit function, two eclipses and a single sinusoid to account for the planet brightness distribution. Limb darkening is fixed to quadratic values obtained with the \texttt{ExoTIC-LD} package \citep{exotic_ld} assuming stellar values from \citet{Burt2024arXiv240514895B}. For NRS1, the star-planet signal is multiplied by a linear trend in time to account for the instrumental drift. We fit for the time of mid-transit $t_0$, orbital inclination $i$, scaled semi-major axis $a/R_\star$, planet-to-star radius ratio $R_{\rm p}/R_\star$, planet-to-star flux ratio $F_{\rm p}/F_\star$, phase curve parameters $C_1$ and $D_1$ following the nomenclature of \citet{cowan2008}, 
an overall offset for each detector $c_0$, and the linear trend for the NRS1 ramp $c_1$. The period of the planet $P$, eccentricity $e$, argument of periastron $\omega$, and quadratic limb darkening coefficients $q_1$ and $q_2$ are fixed. The posterior distribution is explored with nested sampling using 1024 live points using the \texttt{dynesty} package \citep{dynesty}. The best-fit model is shown in Fig.~\ref{fig:phasecurve} and the resulting best-fit values and posteriors are given in Table~\ref{tab:params_nocorr}. A zoomed-in version of the systematics-corrected white light phase curve focusing on the planetary emission signal is shown in Fig.~\ref{fig:phasecurve_zoom}.

We note that the planet radius inferred from our JWST observations is significantly smaller than the latest value reported by \citet{Burt2024arXiv240514895B} using TESS \citep{TESS} and ground-based photometry. Despite using the same stellar parameters as in their work, \citet{Burt2024arXiv240514895B} report a planetary radius of $1.468^{+0.050}_{-0.051}\,R_\oplus$ while we find $1.39\pm0.04\,R_\oplus$ and $1.37\pm0.04\,R_\oplus$ from our fit to the JWST NRS1 and NRS2 data, respectively. Figure~\ref{fig:mass_radius} shows that the change in radius (a 6\% decrement, but 1.37$\sigma$ significant) makes TOI-1685~b even more consistent with having an Earth-like density. This highlights how small changes in the parameter space occupied by small planets --- due to inaccurate stellar parameters, insufficient instrumental precision, stellar activity, inadequate cadence, among others \citep[see, e.g., the discussions in][]{Burt2024arXiv240514895B,AkanaMurphy2024arXiv241102521A} --- have strong consequences on interpreting their internal and atmospheric properties.

\begin{table*}[h!]
    \centering
    \caption{Fitted and derived parameters from the white light phase curve. Prior labels $\mathcal{F}$, $\mathcal{U}$, and $\mathcal{N}$ refer to fixed, uniform (lower bound, upper bound), and normal distributions (mean, standard deviation), respectively.}
\begin{tabular}{l|l|r|r}
\multicolumn{1}{c}{Parameter} & \multicolumn{1}{c}{Prior} & \multicolumn{2}{c}{Posterior}  \\ 
          &                           & \multicolumn{1}{c}{NRS1} & \multicolumn{1}{c}{NRS2} \\ \hline\hline
\multicolumn{4}{c}{Fitted parameters} \\ \hline         
$P$ (d)   & $\mathcal{F}$               & 0.66813924 & 0.66813924 \\ 
$t_0$ (BMJD-60000.5) & $\mathcal{N}$(354.91,0.1)   & $354.912062_{-0.000062}^{+0.000058}$  & $354.912166_{-0.000064}^{+0.000068}$  \\   
$e$            & $\mathcal{F}$               &  0 &  0 \\ 
$\omega$ (deg) & $\mathcal{F}$               & 90 & 90 \\ 
$i$ (deg)      & $\mathcal{N}$(87.17,0.8)    & $86.52_{-0.51}^{+0.54}$   & $86.62_{-0.47}^{+0.53}$  \\     
$a/R_\star$    & $\mathcal{N}$(5.371,0.1)    & $5.462_{-0.088}^{+0.086}$ & $5.460_{-0.081}^{+0.080}$  \\      
$q_1$          & $\mathcal{F}$               & 0.118 & 0.095 \\        
$q_2$          & $\mathcal{F}$               & 0.195 & 0.159 \\       
$R_{\rm p}/R_\star$  & $\mathcal{N}$(0.02751,0.01) & $0.02801\pm0.00019$  & $0.02746\pm0.00021$  \\      
$F_{\rm p}/F_\star$ (ppm)  & $\mathcal{N}$(100,50) & $103_{-5.3}^{+5.0}$  & $122_{-6.7}^{+6.9}$  \\      
$C_1$          & $\mathcal{U}$(0,1)          & $0.359_{-0.018}^{+0.017}$  & $0.4918_{-0.0075}^{+0.0039}$  \\ 
$D_1$          & $\mathcal{U}$(-1,1)         & $-0.520\pm0.033$           & $-0.048\pm0.027$  \\  
$c_0$          & $\mathcal{N}$(1,0.05)       & $0.9999754_{-3.6\mathrm{E}-6}^{+3.8\mathrm{E}-6}$  & $0.9999785\pm4.2\mathrm{E}-6$  \\ 
$c_1$          & $\mathcal{N}$(0,0.01)       & $-0.001808_{-1.0\mathrm{E}-5}^{+1.1\mathrm{E}-5}$  & -  \\ 
\hline
\multicolumn{4}{c}{Derived parameters} \\ \hline         
$F_{\rm p,day}$   (ppm)          & - & $103.1_{-5.3}^{+5.0}$ & $121.7_{-6.7}^{+6.9}$   \\ 
$F_{\rm p,night}$ (ppm)          & - & $ 28.9_{-3.4}^{+3.2}$ & $  1.6_{-1.8}^{+1.0}$   \\ 
$A_{\rm PC}$ (ppm)               & - & $130.3\pm6.0$ & $120.4\pm7.0$           \\ 
$\phi_{\rm PC}$ (\degree E)      & - & $55.3\pm3.0$  & $5.5\pm3.2$    \\ 
$R_{\rm p}$ ($R_\oplus$)         & - & $1.39\pm0.04$ & $1.37\pm0.04$  \\ \hline
\end{tabular}
\label{tab:params_nocorr}
\end{table*}

\subsection{Spectroscopic light curves} \label{subsec:specchan}

\begin{figure*}[t]
    \centering
    \includegraphics[width=0.99\linewidth]{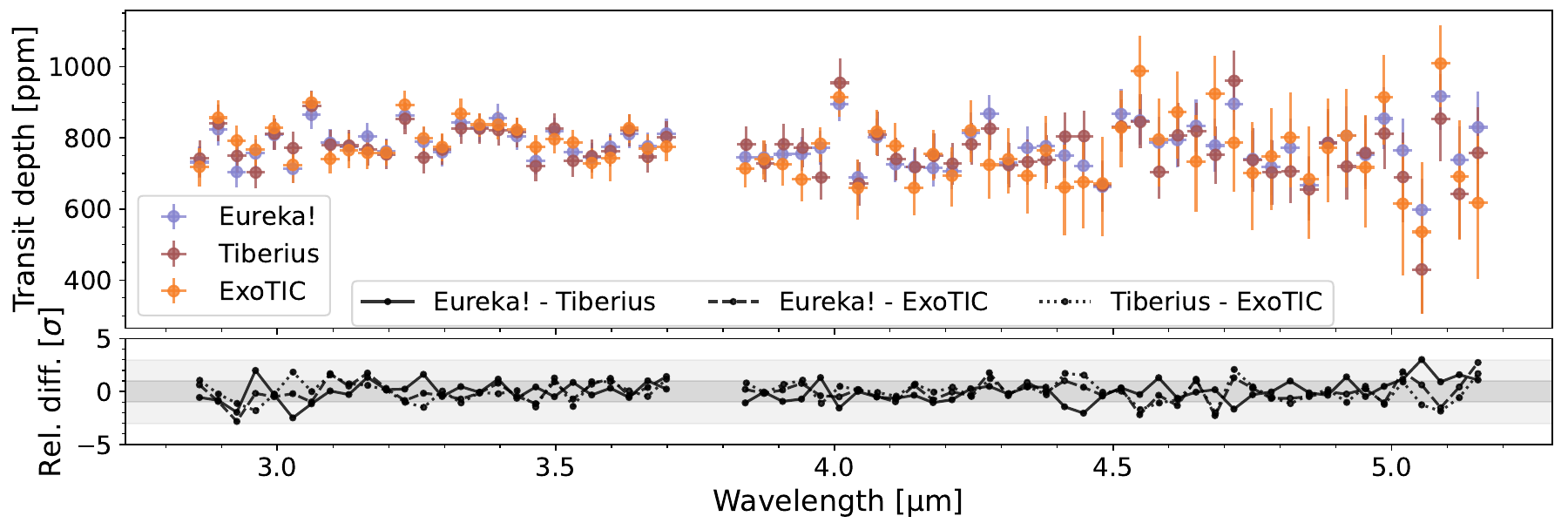}
    \caption{Transmission spectrum of TOI-1685~b using NIRSpec/G395H. Top: Individual transmission spectrum from the \eureka\ (purple), \tiberius\ (red), and \exotic\ (orange) data reductions. Bottom: Difference between each data reduction in units of their standard deviation.}
    \label{fig:transmission_all}
\end{figure*}

\begin{figure*}[t]
    \centering
    \includegraphics[width=0.99\linewidth]{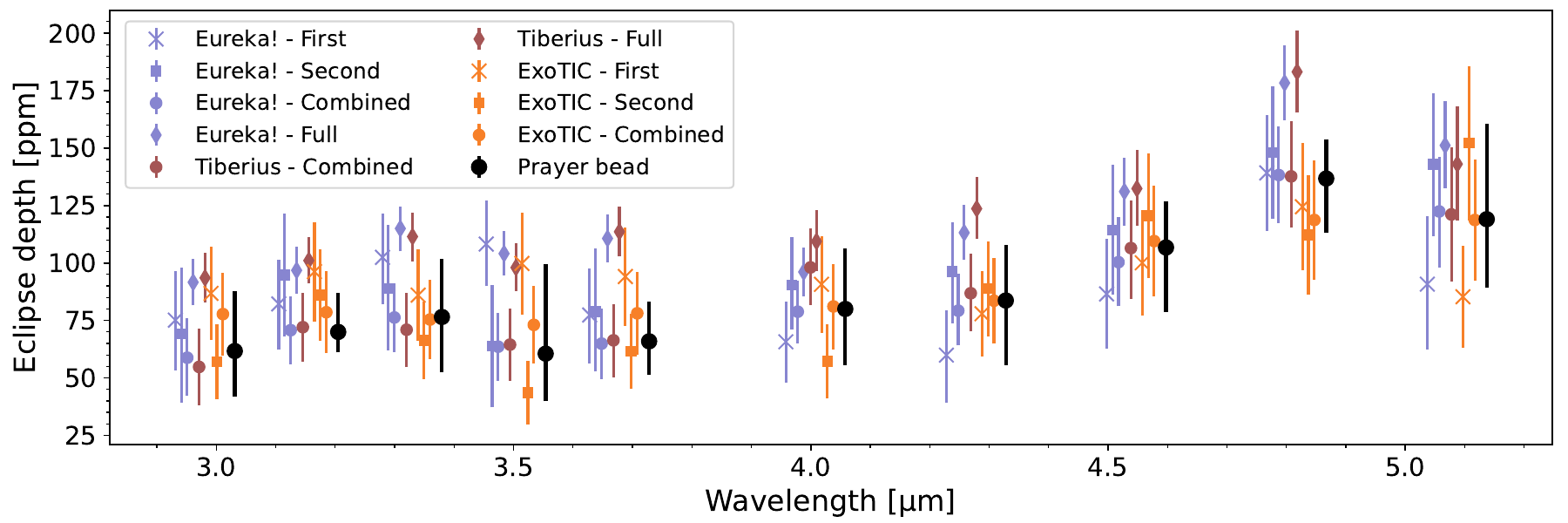}
    \caption{Emission spectrum of TOI-1685~b using NIRSpec/G395H. Different data reductions follow the same color convention as Fig.~\ref{fig:transmission_all}. Different symbols show different parts of the time series used in the fit as described in Sect.~\ref{subsec:specchan}. Black circles represent the final emission spectrum from the prayer-bead analysis carried out in Sect.~\ref{subsubsec:prayerbead}. The different datasets are offset slightly in wavelength to improve visualization. }
    \label{fig:emission_all}
\end{figure*}

For the spectroscopic light curves, we use the same approach as for the white light curves, but in this case, $t_0$, $i$, and $a/R_\star$ are fixed to their best-fit white phase curve model values. Thus, only $R_{\rm p}/R_\star$, $F_{\rm p}/F_\star$, and $c_1$ are free parameters. We use two sets of spectroscopic light curves in our analysis: a higher resolution ($R \sim 87-152$) to study the transmission spectrum and a lower resolution ($R \sim 15-21$) to analyze the emission spectrum. To test the robustness of the transmission and emission spectrum against the baseline model used we carry out different fits that include or exclude different parts of the phase curve. 

\subsubsection{Transmission spectrum} 
For the transmission spectrum, we use two datasets: 1) the full phase curve with the model described in Sect.~\ref{subsec:white} and 2) a transit-only model applied only on the data centered around transit (twice the transit duration before and after the mid-transit time). The posteriors of $R_{\rm p}/R_\star$ per spectroscopic channel are identical in both analyses, thus we show the results from the full phase curve for all three data reductions in Fig.~\ref{fig:transmission_all} and Tables~\ref{tab:transmission_nrs1}~and~\ref{tab:transmission_nrs2}. The consistency between pipelines is remarkable, with the difference between reductions having an RMS between 25 and 32\,ppm for NRS1 and 52-84\,ppm for NRS2. The average transit depth precision per wavelength bin is 40/68\,ppm in NRS1/NRS2 for \eureka, 44/79\,ppm for \texttt{Tiberius}, and 42/119\,ppm for \texttt{ExoTIC-JEDI}. The transit depth precision is slightly better than the 50/80\,ppm per detector predicted by \texttt{PandExo} \citep{pandexo}. Our result shows that previous mismatches between the observed and predicted transit depth precision \citep[e.g.,][]{Alderson2024} may be a consequence of the more complicated noise pattern arising from a low number of groups per integration. Finally, none of the transmission spectra show obvious spectral features --- in line with previous JWST results for rocky exoplanets observed during transit. 

\subsubsection{Emission spectrum} 
For the emission spectrum, we use four datasets: 1) the full phase curve (\textit{``Full"}), 2) an eclipse-only model including only data centered around the two eclipses (\textit{``Combined"}), 3) an eclipse-only model including just the first eclipse (\textit{``First"}), and 4) an eclipse-only model including just the second eclipse (\textit{``Second"}). We apply these analyses to all three data reductions. Some representative emission spectra from each dataset and pipeline are shown in Fig.~\ref{fig:emission_all} and Table~\ref{tab:emission}. As expected, the uncertainties for the first- and second-only eclipses are much larger than the combined or full phase curve time series. It is also important to note how the eclipse depth is 2-3$\sigma$ larger in the full phase curve case compared to fitting the eclipses alone. The more complex model results in deeper eclipses, however, they are too deep compared to the maximum expected given the planet's temperature and indicative of poor treatment of correlated noise present in the complete time series (see below and \S\ref{subsec:albedo}). The average eclipse depth precision is 15/19\,ppm in NRS1/NRS2 when fitting only both eclipses and 10/14\,ppm when fitting the full phase curve. The differences between pipelines are again insignificant, thus we use the \eureka\ reduction for the remainder of analyses in the paper.

\subsection{Correlated noise} \label{subsec:rednoise}

\begin{figure*}
    \centering
    \includegraphics[width=0.99\linewidth]{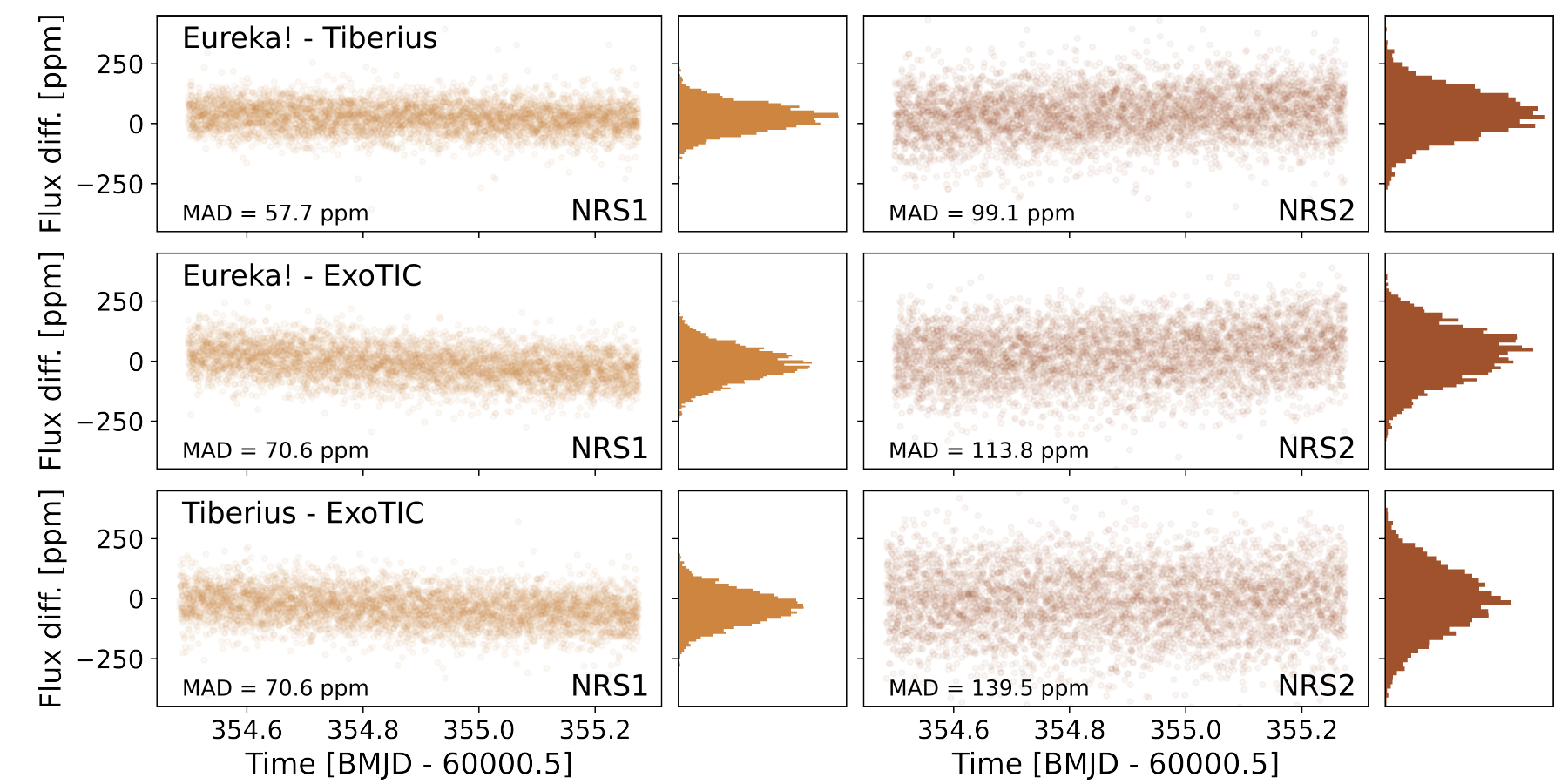}
    \caption{Difference between the white light curves from each data reduction. Each panel shows the median absolute deviation of the data (MAD). Apart from the presence of a small linear trend (-34\,ppm\,d$^{-1}$ in NRS1 and 64\,ppm\,d$^{-1}$ in NRS2), the differences follow a Gaussian distribution indicating that the presence and properties of correlated noise are pipeline-independent. 
    }
    \label{fig:residuals_pc}
\end{figure*}

The residuals in Fig.~\ref{fig:phasecurve} clearly show unaccounted correlated noise in the modeling of the white light phase curve. The RMS of the residuals is 145\,ppm in NRS1 and 170\,ppm in NRS2, which is 2-3 times larger than the measured eclipse depths. This correlated noise has an effect on the emission spectrum accuracy when including or not the complete time series as shown in Fig.~\ref{fig:emission_all} and the accuracy on the amplitude and offset of the phase curve. The origin of this correlated noise appears to be pipeline-independent as shown in Fig.~\ref{fig:residuals_pc}, where the difference between reductions follows normal distributions with very low median absolute deviations (MAD). 

To characterize the noise properties in this data, we ran the residuals between the best-fit model and low-resolution spectroscopic light curves (5 bins) through a Lomb-Scargle \citep{lomb76, scargle82} periodogram, as implemented in \texttt{astropy} \citep{astropy22}. The Lomb-Scargle periodogram is a statistical technique designed to detect periodic signals in any given data set and is regularly used in time series observations. By running the residuals of the spectroscopic light curves through the Lomb-Scargle periodogram, we aim to characterize periodic noise in the data. We provide the results of this test in Figs.~\ref{fig:nrs1_ls} and \ref{fig:nrs2_ls}. 

We find the following results for the \texttt{Eureka!} reduction. For NRS1, we find there is no strong periodic signal at times less than 0.8~hours (Figure~\ref{fig:nrs1_ls}). Several of the spectroscopic channels show periodic signals around $\sim 1$~hour, however, none of the signals between channels are strongly aligned. The spectroscopic channels with $\lambda_\textrm{cen} = 3.10 - 3.45$\,µm have periodic signals at $\sim 2.5$~hours. But again, there is a relative spread in these peaks at $\sim 0.1$~hours. All spectroscopic channels in NRS1 show the strongest periodic signal at $3.7-4.2$~hours. This signal roughly aligns with the timing of the eclipses and transit. The periodicity in NRS2 is significantly different than in NRS1 (Figure~\ref{fig:nrs2_ls}). None of the residuals between spectroscopic channels show any correlation in periodicity. The channel with $\lambda_\textrm{cen} = 3.96$\,µm shows a similarly strong periodic signal at $\sim 4$~hours to the NRS1 data. However, none of the other channels show this level of periodicity.

We performed a battery of tests and analyses described in Appendix~\ref{app:preprocessing} to remove this correlated noise at the data reduction stage, but all our pre-processing attempts were unsuccessful. Therefore, we tried to remove this red noise component using post-processing techniques.

\subsubsection{Principal Component Analysis decorrelation}

To see if changes in the shape of the point-spread function (PSF) are responsible for the correlated noise, we use \texttt{scikit-learn} \citep{Scikit-learn} to perform a principal component analysis (PCA) on the output of \texttt{Eureka!}'s Stage 1, namely the $N_{\rm int} \times N_{\rm rows} \times N_{\rm cols}$ array of fluxes.  For every row, we identify outliers using iterative sigma clipping with a threshold of 5$\sigma$, and interpolate over the bad columns.  We repair NaNs in the same way.  After performing PCA with five components, we inspect the eigenvectors (which are $N_{\rm rows} \times N_{\rm cols}$ images) and manually look for obvious outliers.  Bad pixels are often extremely variable and show up clearly in these eigenvectors.  We identify these outliers and repeat PCA after interpolating them in the input data.  Then, we decorrelate the white light curve against the eigenvalues, which are five 1D time series of length $N_{\rm int}$.  The first eigenvector has the appearance of a negative trace on top of a positive trace, and the corresponding eigenvalues track the y-position of the trace almost perfectly.  We also perform an alternative analysis where we perform PCA without repairing outliers. In both cases, we found no significant reduction in correlated noise upon decorrelating against PCA eigenvalues. We conclude that changes in PSF shape are not responsible for the correlated noise.

Alternatively, we use PCA to attempt to remove the correlated noise at the light curve level. The routine was optimized to preserve the observed transit depth while removing the observed variability not attributed to the phase curve at shorter timescales. We tested this method on both the low- and high-resolution spectroscopic light curves. We use \texttt{scikit-learn} on each set of light curves and visually inspect all of the eigenvectors for each resolution. We removed any eigenvectors that had captured the transit itself, which were the first and third components for each resolution in our case. No other eigenvectors showed a clear transit. 

We ran each eigenvector through a Lomb-Scargle periodogram to determine which vectors captured the short-timescale variability we sought to remove. This periodogram search was less useful for the low-resolution light curves, as we tested various combinations of all 4 available eigenvectors (as the number of eigenvectors must be smaller than the number of light curves themselves, 5) to decorrelate against. For the high-resolution light curves, we selected and tested a combination decorrelating against 1-10 eigenvectors which captured the noise properly. Decorrelating against the eigenvectors reduced the average out-of-transit noise by, on average, 140\,ppm. However, we also found that this decorrelation process changed the transit and eclipse depths, on average, by 80 and 95\,ppm respectively, even though there was no obvious signal from the transit or eclipses in the selected eigenvectors.

\begin{figure}
    \centering
    \includegraphics[width=1.0\linewidth]{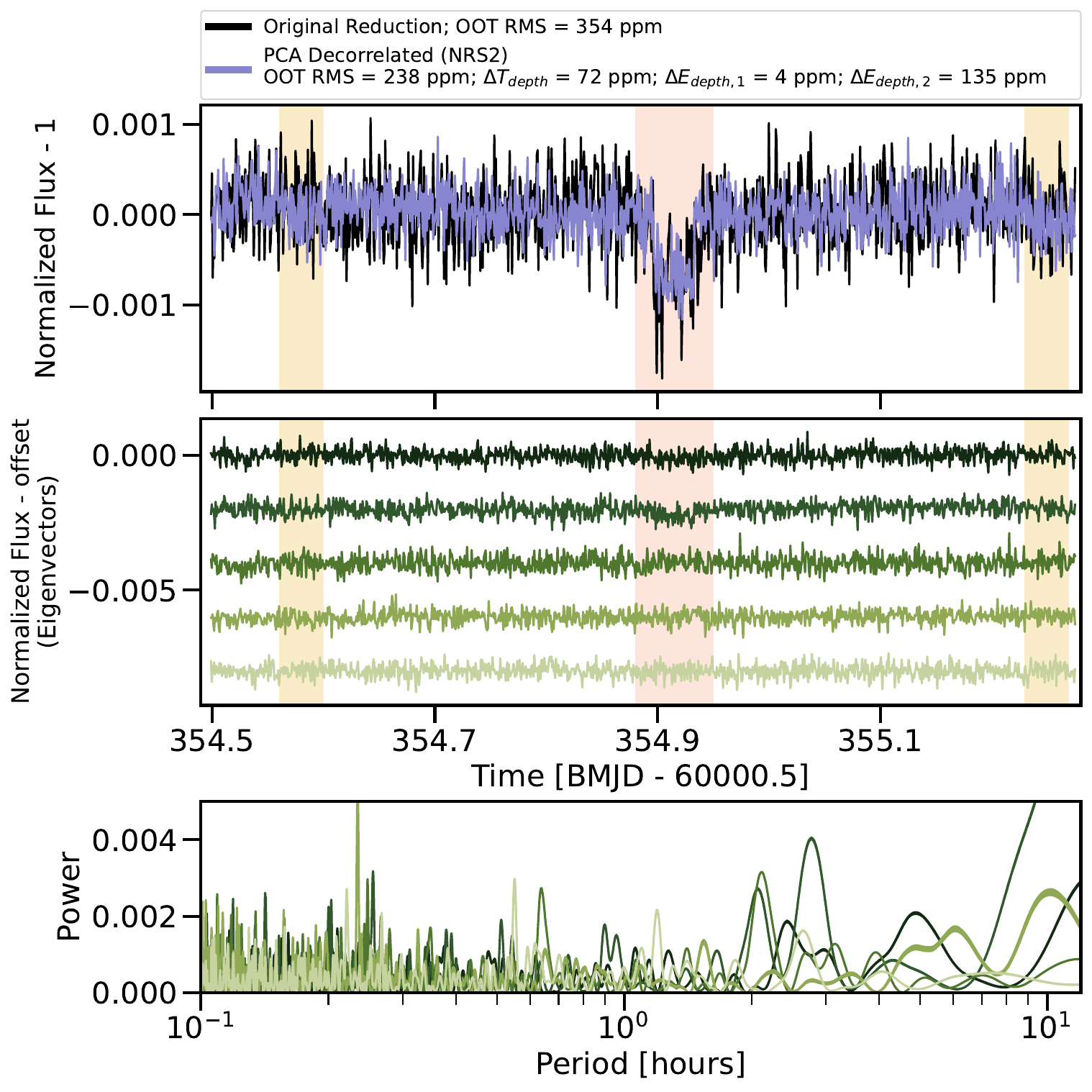}
    \caption{Results of the Independent Component Analysis (ICA) for NRS2. We created 46 independent eigenvectors using the \texttt{FastICA} functionality in \texttt{scikit-learn}. We show the results for a single spectroscopic channel ($\lambda_\textrm{cen} = 4.143 \pm 0.017$). Top: Black is the reduced spectroscopic light curve. Purple is the light curve after it has been decorrelated against the five ``best'' eigenvectors, which are shown in the middle panel. The light curves are binned into 1-minute increments for ease of comparison. Bottom: Results of the Lomb-Scargle periodogram on the five eigenvectors in the middle plot. Colors correspond to the appropriate eigenvector. As noted in the caption, through the ICA decorrelation, we decrease the out-of-transit (OOT) noise from 354 to 238\,ppm. We defined the OOT regions as $t = [(354.63, 354.85), (355.00, 355.20)]$ in units of BMJD - 60000.5. We also note that even though there are little to no residuals of the transit and eclipses in the eigenvectors, the transit (orange-shaded region) and eclipse (yellow-shaded regions) depths change by 72, 4, and 135\,ppm, respectively. The PCA eigenvectors behaved similarly. We conclude that while ICA and PCA can be used to remove similar systematics between light curves, it is best to proceed with caution when applying this technique to chromatic data sets. }
    \label{fig:nrs2_ica}
\end{figure}

\begin{figure}
    \centering
    \includegraphics[width=1.0\linewidth]{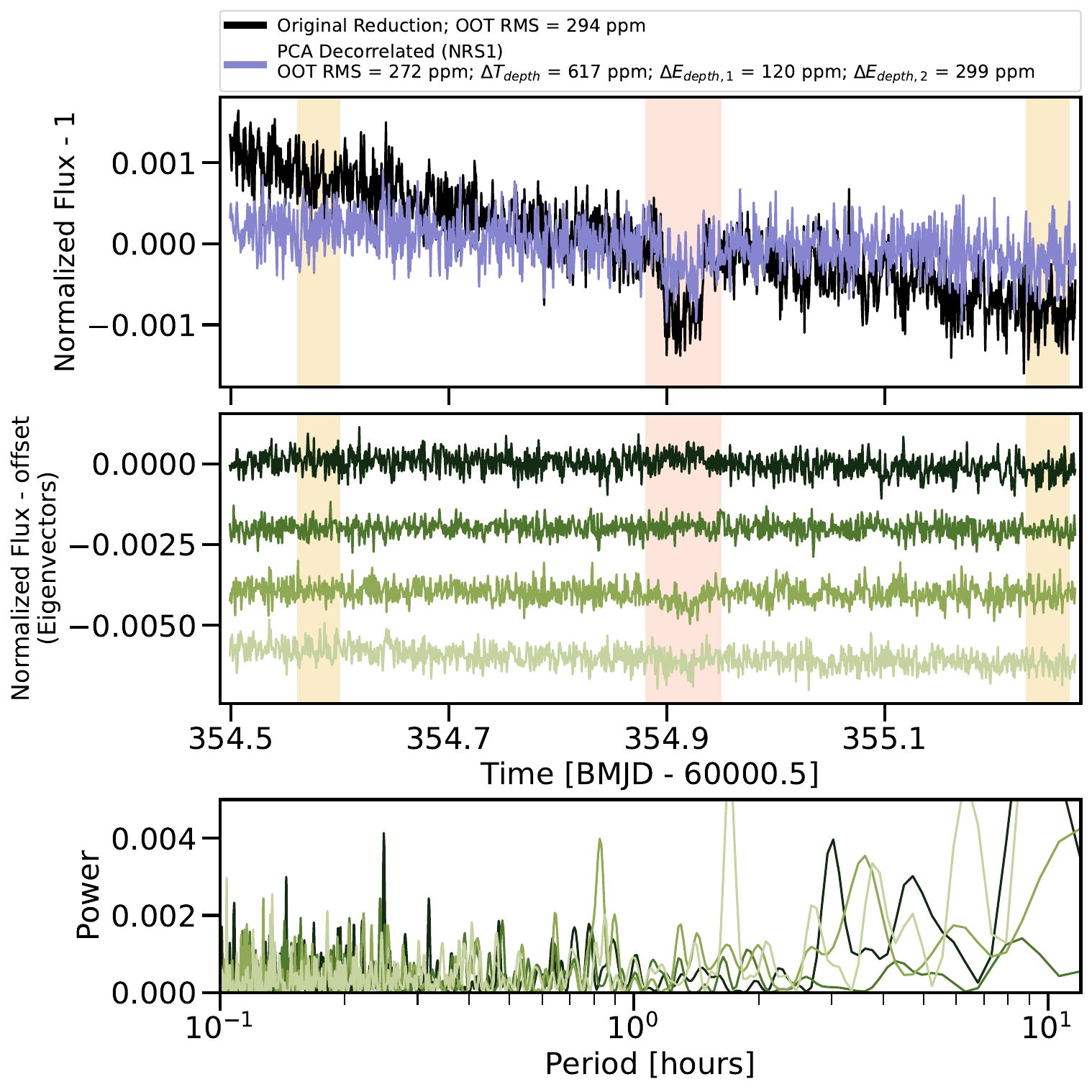}
    \caption{Same as Figure~\ref{fig:nrs2_ica}, but for NRS1. While the ICA eigenvectors are able to successfully remove the linear trend seen in NRS1, nearly every eigenvector showed some residual signal around the transit. This made it challenging to remove the instrumental systematics through this technique without affecting the transit depth, as highlighted here. The eigenvectors for the PCA behaved similarly.}
    \label{fig:nrs1_ica}
\end{figure}

Additionally, we follow a similar methodology, except using Independent Component Analysis (ICA). ICA has the advantage that it is optimized to derive eigenvectors from a dataset that is a mixture of independent sources. In this instance, the independent sources would be the transit, eclipses, stellar noise, and systematic noise. We obtained similar results as our PCA analysis for the ICA analysis. We present the components and results of the ICA as an example for NRS2 in Figure~\ref{fig:nrs2_ica}. We include the equivalent plot for NRS1 in Figure~\ref{fig:nrs1_ica}.

\subsubsection{Prayer-bead} \label{subsubsec:prayerbead}

\begin{table*}
    \centering
    \caption{Median and 1$\sigma$ uncertainties of the fitted parameters from the prayer-bead analysis of the white light phase curve and the low-resolution emission spectrum.}
    \begin{tabular}{c|c|c|c|c|c|c|c}
{Detector} & {{Wavelength}} & {$(R_{\rm p}/R_\star)^2$} & {$F_{\rm p}/F_\star$} & {$C_1$} & {$D_1$} & {{$c_0$}} & {$c_1$} \\
{} & {{(µm)}} & {(ppm)} & {(ppm)} & {} & {} & {} & {}                  \\ \hline\hline

\multicolumn{8}{c}{\textbf{Prayer bead --- White light phase curve}}  \\ \hline
{NRS1} & {2.844 -- 3.715} & {$787.6^{+28.9}_{-33.8}$} & {$103^{+28}_{-27}$} & {$0.36^{+0.21}_{-0.19}$} & {$-0.54^{+0.29}_{-0.34}$} & {$0.99997^{+0.00003}_{- 0.00002}$} & $-0.00181^{+0.00007}_{-0.00006}$  \\
{NRS2} & {3.850 -- 5.172} & {$755.9^{+30.1}_{-29.2}$} & {$119^{+23}_{-18}$} & {$0.49^{+0.13}_{-0.10}$} & {$-0.05^{+0.11}_{-0.10}$} & {$0.99998\pm0.00002$} & - \\ \hline

\multicolumn{8}{c}{\textbf{Prayer bead --- Low-resolution emission spectrum}}  \\ \hline
{\multirow{5}{*}{NRS1}} & {$2.931\pm0.087$} & {\multirow{5}{*}{-}} & {$62^{+26}_{-20}$} & {\multirow{5}{*}{-}} & {\multirow{5}{*}{-}} & {$0.99995^{+0.00001}_{-0.00002}$} & $-0.00181^{+0.00006}_{-0.00003}$  \\
{} & {$3.105\pm0.087$} & {} & {$70^{+17}_{-9}$}  & {} & {} & {$0.99995\pm0.00001$}             & $-0.00234^{+0.00003}_{-0.00002}$ \\
{} & {$3.280\pm0.087$} & {} & {$77^{+25}_{-24}$} & {} & {} & {$0.99995\pm0.00002$}             & $-0.00222^{+0.00008}_{-0.00007}$ \\
{} & {$3.454\pm0.087$} & {} & {$61^{+39}_{-21}$} & {} & {} & {$0.99996^{+0.00001}_{-0.00003}$} & $-0.00145^{+0.00008}_{-0.00009}$ \\
{} & {$3.628\pm0.087$} & {} & {$66^{+17}_{-15}$} & {} & {} & {$0.99995\pm0.00001$}             & $-0.00119^{+0.00003}_{-0.00003}$ \\ \hline
{\multirow{5}{*}{NRS2}} & {$3.958\pm0.135$} & {\multirow{5}{*}{-}} & {$80^{+26}_{-24}$} & {\multirow{5}{*}{-}} & {\multirow{5}{*}{-}} & {$0.99994\pm0.00002$} & {\multirow{5}{*}{-}} \\
{} & {$4.228\pm0.135$} & {} & {$84^{+24}_{-28}$}  & {} & {} & {$0.99994\pm0.00002$}             & {} \\
{} & {$4.498\pm0.135$} & {} & {$106^{+20}_{-28}$} & {} & {} & {$0.99992^{+0.00002}_{-0.00001}$} & {} \\
{} & {$4.767\pm0.135$} & {} & {$137^{+17}_{-23}$} & {} & {} & {$0.99990^{+0.00002}_{-0.00001}$} & {} \\
{} & {$5.037\pm0.135$} & {} & {$119^{+41}_{-30}$} & {} & {} & {$0.99992^{+0.00002}_{-0.00003}$} & {} \\ \hline
    \end{tabular}\label{tab:lc_best}
\end{table*}

Given the unsuccessful attempts at removing the correlated noise using pre- or post-processing techniques, we must model the data in a robust and conservative way that does not underestimate the uncertainties in the fitted and derived parameters. To accomplish this,  we perform a prayer-bead analysis (\citealt{cowan_thermal_2012}, \S 4.2) on the white-light phase curve and the low-resolution emission spectrum (as the results from the transmission spectrum are unaffected by accounting or not for the correlated noise given the high signal-to-noise of the transit depth measurement) obtained with the \texttt{Eureka!} reduction. This technique allows us to estimate the model parameter uncertainties better than using the scatter in the residuals of the best-fit model. In summary, the prayer-bead analysis maintains the relative ordering of the residuals and simply shifts them all by the same amount (wrapping around the start/end of the data), so that correlated noise present in the residuals is preserved. By definition, the prayer bead has as many iterations as there are data points, which in our case is 4370 for the final light curve. For both the low-resolution emission spectrum and the white light phase curve, we fit the light curves using the priors from \ref{subsec:white} and \ref{subsec:specchan}, shift the residuals by one exposure time, add it back to the best-fit model to produce a new light curve, and finally fit the new light curve using the same model and priors. For the white light phase curve, we use a Markov Chain Monte Carlo (MCMC) optimization process using the package \texttt{emcee} \citep{mackey13} to have more reliable uncertainties on the parameters. We used 100 walkers and 1000 steps for the sampling. For the low-resolution emission spectrum, given the long runtimes of performing a prayer-bead analysis on 10 spectroscopic channels (10 $\times$ 4370 model fits), we obtain the best-fit model by simply minimizing the log-likelihood with \texttt{scipy.optimize.minimize} using the modified Powell algorithm \citep{minimization_Powell}. The median and 1$\sigma$ uncertainties for all fitted parameters are reported in Table~\ref{tab:lc_best}.

For the white-light prayer-bead analysis, we reproduce the posterior by generating a Gaussian distribution with 1000 points using the median and $\pm 1\sigma$ values from each iterated phase curve MCMC fitting, resulting in $4370 \times 1000$ samples. Figures~\ref{fig:corner_prayerbead_nrs1}~and~\ref{fig:corner_prayerbead_nrs2} show the posterior distributions of the fitted parameters to the white light phase curve from NRS1 and NRS2, respectively. The median and 1$\sigma$ uncertainties are reported in Table~\ref{tab:lc_best}. As discussed in \citet{cowan_thermal_2012}, the prayer bead is a very conservative technique that provides large errorbars, especially in the phase variation parameters, making us confident that we are not underestimating the true uncertainty of the derived parameters from the JWST observations of this planet under the presence of misunderstood systematics. Therefore, we use the results from the prayer-bead analysis in the analyses of the emission spectrum and white-light phase curve parameters shown in Table~\ref{tab:lc_best} for the rest of the paper.

\section{Results and Discussion} \label{sec:discussion}

\subsection{Constraints from the transmission spectrum}



\subsubsection{Atmospheric retrievals} \label{subsec:retrievals_tra}

\begin{figure*}
    \centering
    \includegraphics[width=1.0\linewidth]{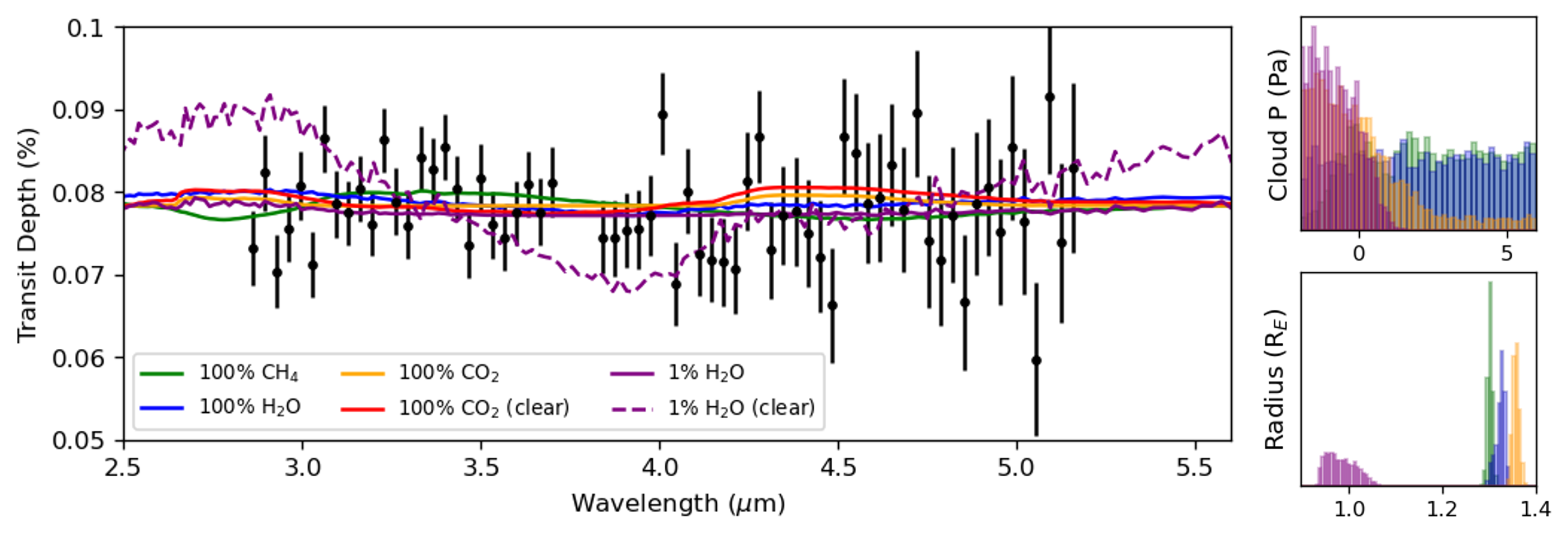}
    \caption{Summary of transmission retrievals with TauREx. Left: Best-fit spectra and transit observations (black data points) for TOI-1685\,b. Right: Probability density for the four runs including clouds. Our retrieval exploration rejects the clear primary atmosphere case. For the pure CO$_2$ case, clouds can help reduce the 4.5$\mu$m absorption feature, but this is not a statistically significant result (see Bayesian Evidence in Table \ref{tab:taurex_transits}).}
    \label{fig:TauREx_tra}
\end{figure*}

To comprehensively analyze the transmission spectrum, we utilize the retrieval framework TauREx3 \citep{TauREx,TauREx3p1}. Given the apparent featureless nature of the spectrum, we explore simple atmospheres only (with a single radiatively active molecule, i.e., one species). The atmosphere of TOI-1685~b is modeled using a 1D plane-parallel atmosphere with 70 layers spanning 10\,bar to 1\,µbar linearly spaced in log-pressure. The abundance of the possible species -- H$_2$O, CO$_2$, and CH$_4$ -- are therefore fixed to either 100\% (hypothesis for a secondary atmosphere) or 1\% (hypothesis for a primary atmosphere with trace species). A fully opaque cloud deck is added to explore the possibility of muted features. We retrieve the planetary radius and cloud top pressure. We also explore models without clouds (clear case), hence only the planetary radius is retrieved, to rule out scenarios producing large spectral features. A summarizing figure containing the spectra and probability density is shown in Fig.~\ref{fig:TauREx_tra}. 

With these simple retrievals, we conclude that TOI-1685\,b cannot have a clear H$_2$-dominated atmosphere. All the secondary atmospheres we tested are consistent with the data: the spectral features resulting from a 100\% H$_2$O, 100\% CO$_2$, and 100\% CH$_4$ are small enough to be within the uncertainties of the data, even in the case of a clear atmosphere. We note that the pure CH$_4$ atmospheric model has marginally higher Bayesian evidence than a flat line (see Table \ref{tab:taurex_transits}). However, a pure methane atmosphere would be unexpected from theory, given its ready photodissociation likely leading to significantly heavier hydrocarbon species and ultimately a thick atmospheric haze that would remove the presence of significant CH$_4$ from the atmospheric spectrum itself \citep[e.g.,][]{Bergin2023ApJ...949L..17B}.

\begin{table*}[]
\centering
\caption{TauREx3 retrieval interpretations of the transit data. The observations are consistent with a flat line. Using the Bayesian Evidence from the \textsc{MultiNest} runs, noted ln(E), only the clear primary atmosphere scenario can be ruled out confidently. }
\begin{tabular}{|c|c|c|}
 \hline
 Model & ln(E) & Remarks  \\ \hline 
 No chemistry (flat line)   & 546.7 & Consistent with data. \\ 
 Clear 99\% H$_2$ + 1\%H$_2$O & 505.5 & A clear H$_2$-dominated atmosphere is rejected ($\Delta$ln(E) $\sim$ 41). \\ 
 Cloudy 99\% H$_2$ + 1\%H$_2$O & 545.1 & A cloudy H$_2$-rich atmosphere is possible, but it implies $R_{\rm p} < 1\,R_\oplus$ and P$_\mathrm{c} < 1$\,Pa. \\ 
 Cloudy 100\% H$_2$O & 547.5 & A pure H$_2$O atmosphere with or without clouds is possible. \\ 
 Cloudy 100\% CH$_4$ & 549.6 & A pure CH$_4$ is moderately favored ($\Delta$ln(E) $>$ 3), but physically unlikely. \\
 Cloudy 100\% CO$_2$ & 546.6 & A pure CO$_2$ atmosphere is possible, but clouds posteriors imply P$_c$ $<$ 10\,Pa. \\
 Clear 100\% CO$_2$  & 544.9 & There is poor evidence for clouds in the CO$_2$ case ($\Delta$ln(E) $<$ 3). \\ \hline 
\end{tabular}    \label{tab:taurex_transits}
\end{table*}

\subsubsection{Non-physical models} \label{subsec:unphys_tra}

\begin{table}[h!]
\centering
\caption{Results from non-physical model fits to the transmission spectrum.}
\begin{tabular}{l|ccccc}
\textbf{Model Type} & $\log Z$  & $\chi^2/N$    & Free parameters       \\ \hline
Flat line           & -39.7     & 1.09  & 1     \\
Offset NRS1/NRS2    & -41.2     & 1.05  & 2     \\
Slope               & -42.3     & 1.08  & 2     \\
Gaussian Feature    & -39.9     & 1.08  & 4     \\ \hline \hline
\end{tabular}
\label{tab:gaussian_mods}
\end{table}

To explore potential atmospheric conditions not explicitly included in our atmospheric retrievals, we tested a series of non-physical models against the transmission spectrum using a nested sampling algorithm using \texttt{MLFriends} \citep{Buchner2019} within \texttt{Ultranest} \citep{Buchner2021}. This agnostic method of feature detection follows previous rocky atmospheric characterization with JWST \citep[e.g.,][]{MoranStevenson2023,Alderson2023,Wallack2024}. These agnostic non-physical models include a 1) flat line, 2) an offset between the NRS1 and NRS2 detectors, 3) a sloped line, and 4) a Gaussian-shaped spectral feature. This last case fits for the amplitude, width, and midpoint of a putative Gaussian-shaped feature. We compute the log-likelihood of each model and then a Bayes factor \citep{Trotta2008} between them. 

The results of this nested sampling comparison are shown in Table \ref{tab:gaussian_mods}. The flat line has the highest log-likelihood of any model tested regardless of data reduction. All non-physical models have very similar $\chi^2/N$.  The increase in complexity of the non-flat models never results in a correspondingly significant increase in the goodness of fit. Thus in all cases, a flat line is preferred against any of the other potential Gaussian, offset, or slope models.

\subsection{Constraints from the dayside emission spectrum and white light phase variations}

\subsubsection{Albedo and heat recirculation efficiency} \label{subsec:albedo}

\begin{figure}
    \centering
    \includegraphics[width=1.0\linewidth]{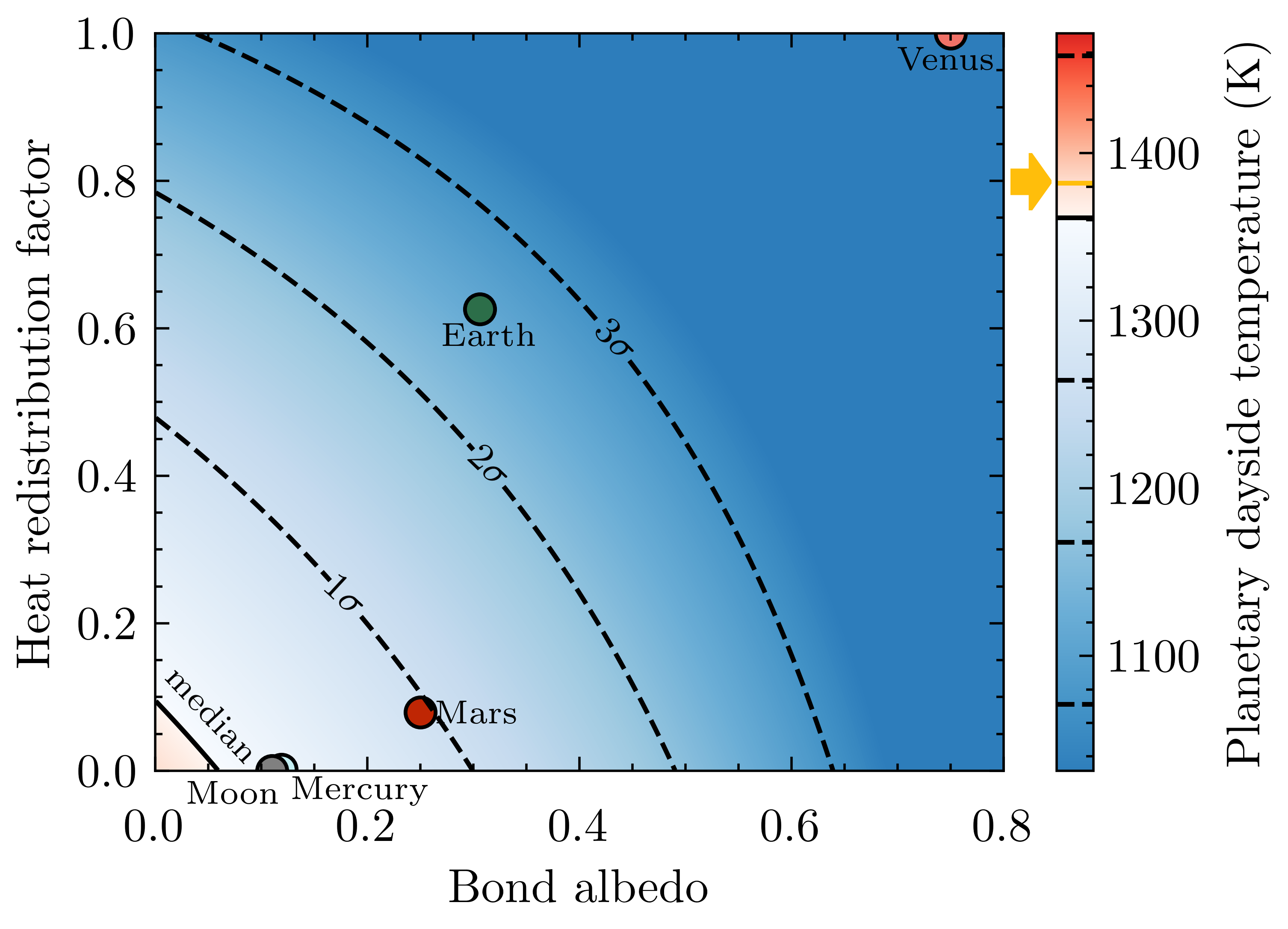}
    \caption{Contour plot showing the planetary dayside temperature of TOI-1685\,b as a function of Bond albedo and heat redistribution factor ($\varepsilon$) compared to estimates for Solar System bodies from \citet{xue24,mansfield24}. The yellow arrow indicates T$_{p,dayside}$ for a blackbody (assuming A$_B$ = 0 and no heat recirculation.) The large uncertainties on the derived T$_{p,dayside}$ allow a wide range of interpretation, but the results are 1-$\sigma$ consistent with the (almost) airless Moon, Mercury, and Mars.}
    \label{fig:albedo_heatredist}
\end{figure}

We derive the dayside and nightside brightness temperatures of TOI-1685~b from the white light phase curve and the low-resolution emission spectrum following a similar method as in \citet{zhang24}. From these, we can estimate the effective albedo $A_{\rm eff}$ and heat recirculation efficiency $\varepsilon$ using the definitions from \citet{Cowan2011ApJ...726...82C}.

From the white light phase curve, we compute the dayside and nightside brightness temperatures by multiplying the dayside $F_{\rm p,day}/F_\star$ and nightside $F_{\rm p,night}/F_\star = F_{\rm p,day}/F_\star \times (1-2C)$ planet-to-star flux ratio samples from the prayer-bead analysis by a stellar spectrum obtained from interpolating the PHOENIX grid \citep{phoenix} using the python package \texttt{pysynphot} \citep{pysynphot} to the stellar photospheric parameters reported in \citet{Burt2024arXiv240514895B}. In our calculation, we account for uncertainties in the planet radius (5\%, \citealt{Burt2024arXiv240514895B}) and systematic deviations between the true stellar spectrum in the NIRSpec passband and the PHOENIX model (1\%, consistent with the findings by \citealt{zhang24} for MIRI LRS). As in the rest of our analyses, we model NRS1 and NRS2 independently. 

We obtain $T_{\rm p,day} = 1520_{-140}^{+140}\,\mathrm{K}$ and $T_{\rm p,night} = 1100_{-1100}^{+210}\,\mathrm{K}$ for NRS1, and $T_{\rm p,day} = 1360_{-100}^{+100}\,\mathrm{K}$ and $T_{\rm p,night} = 550_{-550}^{+300}\,\mathrm{K}$ for NRS2. For comparison, the dayside brightness temperature of a zero Bond albedo, zero heat recirculation blackbody would have $T_{\rm p,max} \sim 1390\,\mathrm{K}$. Defining the temperature scaling ratio $\mathcal{R} \equiv T_{\rm p,day}/T_{\rm p,max}$, we obtain $\mathcal{R} = 1.10\pm0.10$ and $0.98\pm0.07$ for NRS1 and NRS2, respectively. The effective albedo is then $A_{\rm eff} = -1.0_{-0.9}^{+1.0}$ and $0.0_{-0.4}^{+0.3}$ for NRS1 and NRS2, respectively. The heat recirculation efficiency is $\varepsilon = 0.44_{-0.44}^{+0.36}$ and $0.06_{-0.06}^{+0.24}$ for NRS1 and NRS2, respectively. 

We note that the results from NRS1 have much larger uncertainties than those from NRS2, even suggesting a negative albedo, which would suggest the planet emits more energy than it receives. On the one hand, we find that the correlated noise is stronger in NRS1 compared to NRS2 and it has a longer periodicity of approximately 4\,hours. This timescale is one quarter of the total orbit of the planet and affects the determination of the phase curve parameters modeled with a sinusoid. On the other hand, the negative linear trend seen in the NRS1 detector introduces a degeneracy with the phase curve parameters. Therefore, although we report values for both detectors in Table~\ref{tab:params_nocorr}, the parameters derived from NRS2 data are more reliable. Figure~\ref{fig:albedo_heatredist} shows the dayside brightness temperature from NRS2 as a function of Bond albedo and heat redistribution as computed in \citet{xue24}. The large uncertainties in $T_{\rm p,day}$ allow a broad range of solutions, but the median and 1$\sigma$ values on the Bond albedo and heat redistribution efficiency are consistent with the parameters of the (nearly-)airless bodies of the solar system (Moon, Mercury, and Mars).

For completeness and to check the reliability of our white light phase curve results (which is the dataset most affected by unaccounted systematics), we also compute the brightness temperature of TOI-1685~b from the low-resolution emission spectrum.  We use nested sampling code \citep{dynesty} to fit the spectrum, with the free parameters being the ratio $\mathcal{R}$, the stellar parameters ($T_{\rm eff}$, $\log g$, [M/H]), the transit parameters $R_{\rm p}/R_\star$ and $a/R_\star$, and a multiplier that accounts for theoretical uncertainties in the PHOENIX stellar spectrum. $\mathcal{R}$ is given a uniform prior, while the stellar and transit parameters are given Gaussian priors with means and standard deviations equal to the measured values and their errors as in \citet{Burt2024arXiv240514895B} and Table~\ref{tab:lc_best}, respectively. The multiplier is given a Gaussian prior with a mean of 1 and standard deviation of 0.03, reflecting an assumed 3\% uncertainty. We find $\mathcal{R} = 0.94\pm0.04$ for both NRS1 and NRS2. Although the temperature scaling ratio is still consistent within 1$\sigma$ with the NRS2 result from the white light phase curve, the difference is noticeable and consistent with recent results in the literature \citep{xue24,mansfield24}.

\subsubsection{Forward models} \label{subsec:emis_forward}

\begin{figure*}
    \centering
    \includegraphics[width=1.0\linewidth]{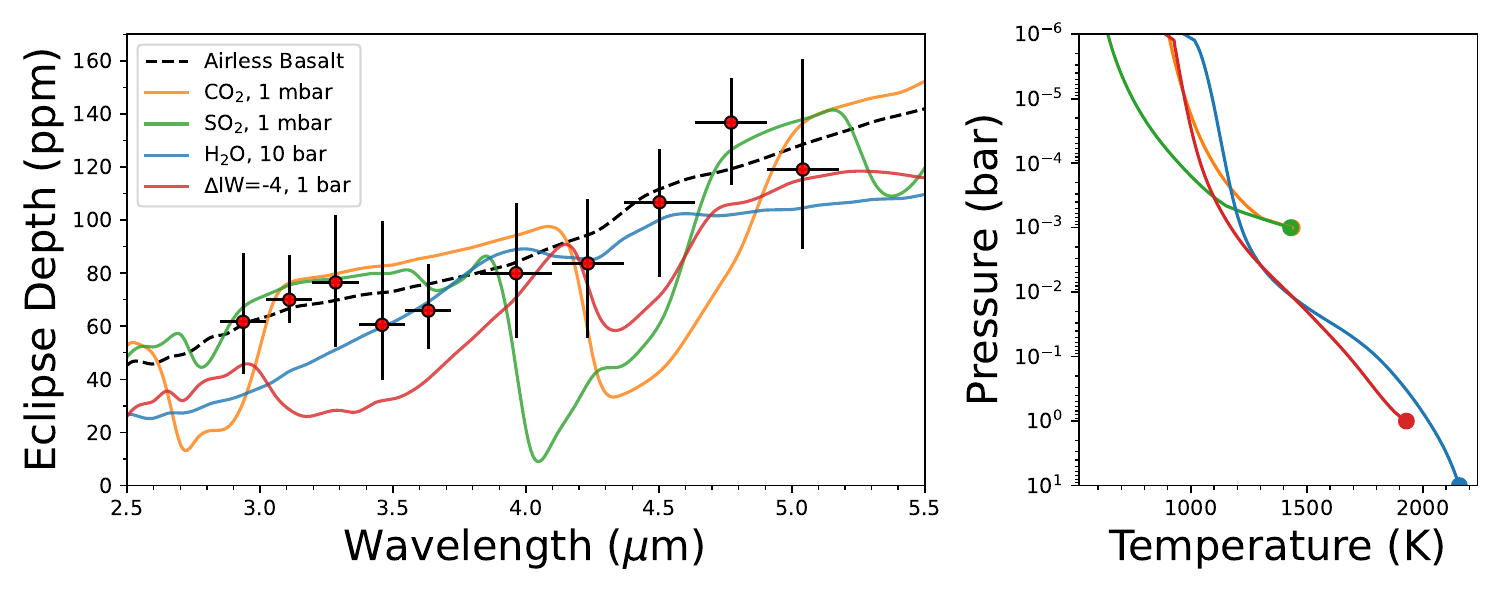}
    \caption{Example poor-fitting emission models of atmospheres and their associated temperature-pressure profiles using \texttt{HELIOS}. In all tested atmospheres, the airless basaltic model is preferred.}
    \label{fig:atmospheres_emission}
\end{figure*}

\begin{table}[]
    \centering
    \caption{Consistency metrics of atmosphere forward models in emission.}
    \begin{tabular}{c|c|c|c|c}
         Model & $P_{atm}$ & $\chi^2\ (N=10)$ & $\Delta\ln{Z}$ & Consistent? \\  \hline \hline
         Airless & -- & 1.8 & 0 & \textbf{Y} \\ \hline
         CO$_2$ &  10\,µbar &  4.6 & $-1.4$ & \textbf{Y} \\
         CO$_2$ & 100\,µbar & 10.2 & $-4.1$ & N \\  \hline
         SO$_2$ & 100\,µbar &  3.5 & $-0.8$ & \textbf{Y} \\ 
         SO$_2$ &   1\,mbar & 12.0 & $-5.0$ & N \\ \hline
         H$_2$O &   1\,bar  &  5.9 & $-2.1$ & \textbf{Y} \\ 
         H$_2$O &  10\,bar  & 10.3 & $-3.6$ & N \\ \hline
         $\Delta\mathrm{IW}=-4$ & 100\,µbar &  2.4 & $-0.3$ & \textbf{Y} \\ 
         $\Delta\mathrm{IW}=-4$ &   1\,mbar &  9.1 & $-3.4$ & N \\ \hline
         $\Delta\mathrm{IW}=0$  & 100\,µbar &  6.2 & $-2.1$ & \textbf{Y} \\ 
         $\Delta\mathrm{IW}=0$  &   1\,mbar & 13.4 & $-5.8$ & N \\ \hline
         $\Delta\mathrm{IW}=+4$ &  10\,µbar &  3.8 & $-1.0$ & \textbf{Y} \\
         $\Delta\mathrm{IW}=+4$ & 100\,µbar & 10.5 & $-4.3$ & N \\ \hline
    \end{tabular} \label{tab:atm_rejection}
\end{table}

We investigate the possibility of thin, possibly tenuous atmospheres by forward modeling emission spectra using the radiative transfer code \texttt{HELIOS 3.0} \citep{helios1,helios2,helios3,whittaker22}.  We follow the approach used by \citet{zhang24} to investigate what atmospheres can be confidently rejected by the data.  First, we use \texttt{HELIOS} to calculate forward models for a variety of atmospheric compositions and thicknesses.  We consider cloud-free pure CO$_2$, SO$_2$, and H$_2$O atmospheres, as well as the compositions expected to be outgassed from a magma ocean at the oxidation states $\Delta\mathrm{IW}=-4$, $\Delta\mathrm{IW}=0$, and $\Delta\mathrm{IW}=+4$ from \citet{gaillard22}. Compositions are as follows: $\Delta\mathrm{IW}=-4$ (reduced case):  56\% H$_2$, 41\% CO, 1.4\% CH$_4$, 0.89\% N$_2$, 0.52\% H$_2$O, 0.1\% CO$_2$, and H$_2$S; $\Delta\mathrm{IW}=0$ is 74\% CO, 19\% CO$_2$, 3.3\% 0.89\% N$_2$, 1.9\% H$_2$, 1.8\% H$_2$O, and 0.53\% H$_2$S; and $\Delta\mathrm{IW}=+4$ (oxidized case) is 58\% CO$_2$, 35\% SO$_2$, 2.6\% N$_2$, 2.3\% CO, 0.9\% H$_2$O, and 90 ppm H$_2$.  While TOI-1685~b likely does not possess a global magma ocean, these compositions are representative of outgassed atmospheres expected at different mantle oxidation states.  Deviating from \citet{zhang24}, we use the scaling law derived in \citet{koll22} implemented in \texttt{HELIOS 3.0} to estimate the heat redistribution of each atmospheric scenario.  
We use \texttt{dynesty} nested sampling with a single scaling factor to account for uncertainties in the emitting surface area (5.6\%) and semi-major axis (3.0\%, \citealt{Burt2024arXiv240514895B}), added in quadrature (6.4\%). We use this to calculate the log evidence ($\ln(Z)$) of each forward model with respect to the \eureka prayer-bead emission spectrum. The evidence of our forward models is then compared to that of an airless basaltic surface. 

Our emission spectrum covering from 2.84--5.17\,µm is particularly sensitive to SO$_2$ and CO$_2$ spectral features. Therefore, the lack of features can rule out exceedingly thin (0.1--1\,mbar) clear atmospheres for these species with our modeled temperature-pressure profiles.  Below this threshold, spectral features begin to disappear almost entirely, and the models largely resemble an airless basalt.  Pure H$_2$O atmospheres are difficult to rule out with certainty due to the lack of prominent spectral features. However, the lack of observed heat redistribution can rule out a 10\,bar H$_2$O atmosphere, in which case the heat redistribution would be much higher.


\subsubsection{Atmospheric retrievals} \label{subsec:emis_retrievals}

\begin{figure*}
    \centering
    \includegraphics[width=1.0\linewidth]{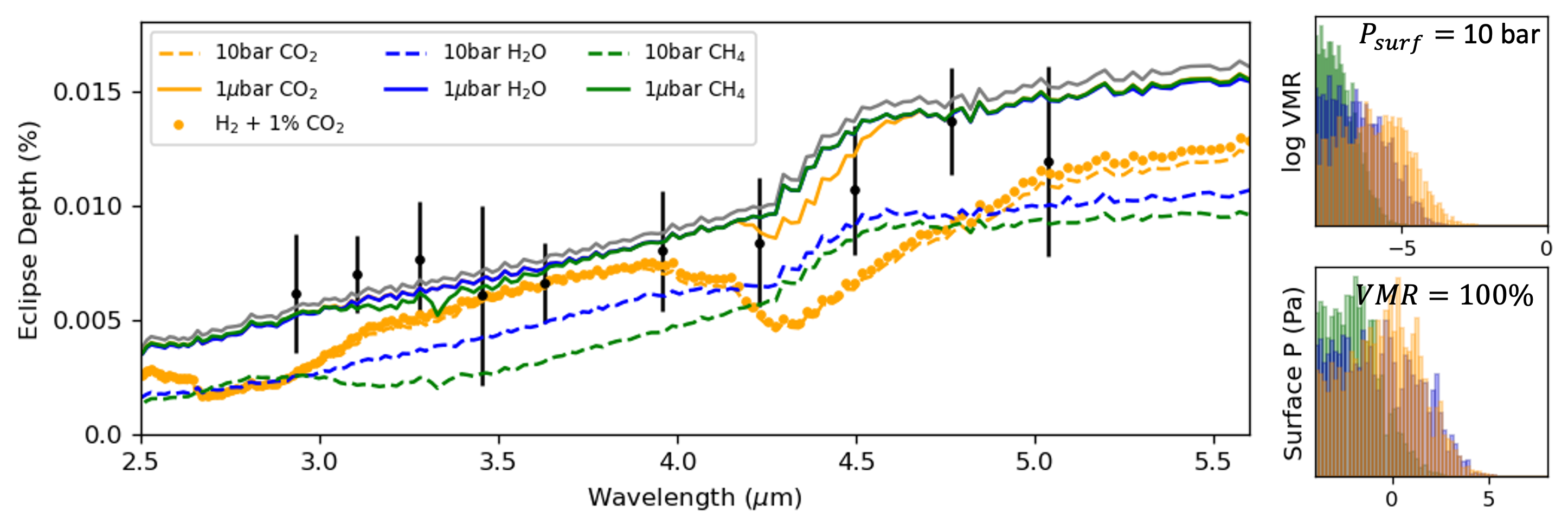}
    \caption{Summary of emission retrievals with TauREx3. Left: Best-fit spectra and transit observations (black data points) for TOI-1685\,b. Right: Probability density for runs where the surface pressure is fixed (top right) or the VMR is fixed to 100\,\% (bottom right). Our retrieval exploration indicates that if the planet hosts a thick atmosphere, only low mixing ratios of H$_2$O, CO$_2$, and CH$_4$ are allowed, which contradicts the results on the transit data. If TOI-1685\,b hosts a pure atmosphere, only µbar levels are allowed. The combination of transit and eclipse information suggests a low-pressure atmosphere or no atmosphere at all.}
    \label{fig:TauREx_em}
\end{figure*}

\begin{figure*}
    \centering
    \includegraphics[width=1.0\linewidth]{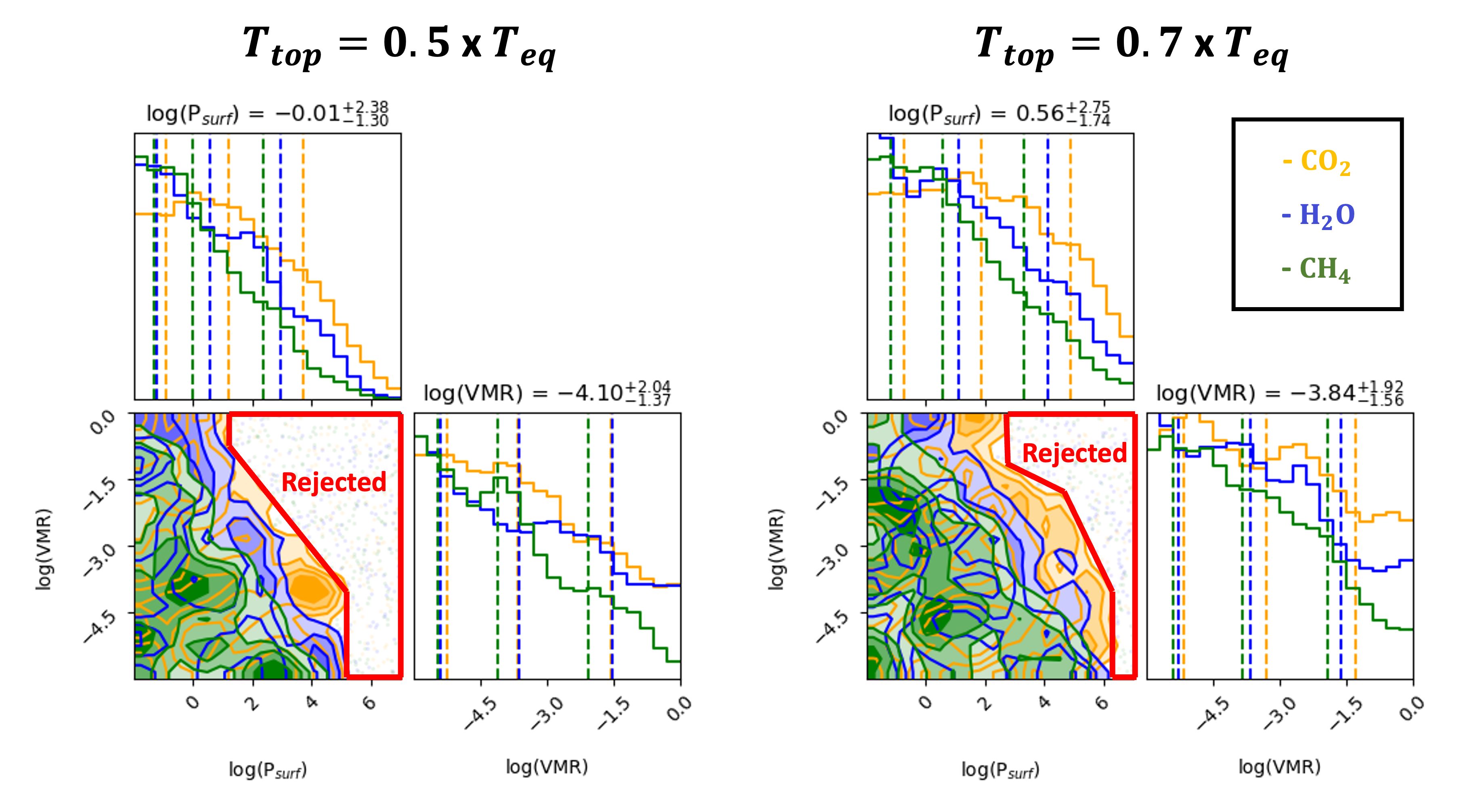}
    \caption{Corner plot for emission results with TauREx3. The forward models of those retrievals have a fixed radius and temperature gradient, allowing to reject some part of the parameter space. Since the eclipse spectra have large uncertainties and do not contain obvious spectral features (i.e., the eclipse data is consistent with a blackbody emission), the thermal gradient information cannot be inferred from the data and must be assumed. Overall, independently from the main molecular species, the regions with high surface pressures (P $>$ 10\,bar) are rejected, but this result might depend on the thermal gradient assumption.}
    \label{fig:TauREx_corner}
\end{figure*}

For an alternate way of interpreting the eclipse spectrum, we follow a similar procedure as in Sect.~\ref{subsec:retrievals_tra} and rely on single radiatively active species retrievals. At low SNR, eclipses are much more challenging to interpret than transits since the emission spectrum is a complex convolution of the planetary radius, the surface pressure, the thermal gradient, and the chemical composition. To conduct this rejection experiment, we therefore reduce the degrees of freedom in our retrievals to the maximum. We fix the planetary radius to median values retrieved from the transit retrievals (i.e., $R_p \in \{1.327, 1.358, 1.303\}\,R_\oplus$ for respectively the H$_2$O, the CO$_2$, and the CH$_4$ cases) to maintain consistency and since the transit data is much more sensitive to the radius information. We also assume a thermal profile. The thermal profile is defined by the surface temperature (fixed to $T_\mathrm{bot} = 1.2\,\times\,T_\mathrm{eq}$, where $T_\mathrm{eq}$ is the equilibrium temperature) and we test different thermal gradient via the top temperature ($T_\mathrm{top} \in \{0.7, 0.5\}\,\times\,T_\mathrm{eq}$). We explore the same molecules as in the transit case (i.e., H$_2$O, CO$_2$, and CH$_4$) and retrieve the surface pressure fixed ($P_\mathrm{surf}$) and the abundance of the main molecule (Volume Mixing Ratio, VMR). The rest of the atmosphere is filled with a H$_2$/He mixture at solar ratio. Although these two scenarios may not be representative of TOI-1685~b's atmosphere, they allow us to explore the extent of possible solutions. We show in Fig.~\ref{fig:TauREx_em} forward models and posterior probability density for those cases, as well as the posterior distributions in Fig.~\ref{fig:TauREx_corner}.

The eclipse spectrum does not show any spectral features (i.e., consistent with blackbody planetary emission). Overall, while our retrieval investigations of the eclipse spectrum are idealized (i.e., simple atmospheres), they appear to rule out thick atmosphere cases with high mean molecular weight. However, as shown by the red regions in the corner plots of Fig.~\ref{fig:TauREx_corner}, the upper limit on the surface pressure is highly dependent on the chemical composition and the atmospheric thermal gradient. As such, it is difficult to provide a strict upper limit on the atmospheric surface pressure.    

Taken together, our retrieval analysis of the emission spectrum of TOI-1685~b is compatible with a very low-pressure atmosphere. For a high surface pressure atmosphere to exist, a thick H$_2$-dominated atmosphere and a small thermal gradient would be required, which is inconsistent with the data (the transmission spectrum and the planet's bulk density) and unlikely from a physical point of view (given the high equilibrium temperature of the planet and the possibility of atmospheric mass-loss in the past). Combined with a lack of nightside emission detection (i.e., no energy redistribution), we conclude that TOI-1685~b is likely a bare rock with no significant atmosphere. 

\subsubsection{Surface Properties} \label{subsec:surface}

\begin{figure}
    \centering
    \includegraphics[width=0.99\linewidth]{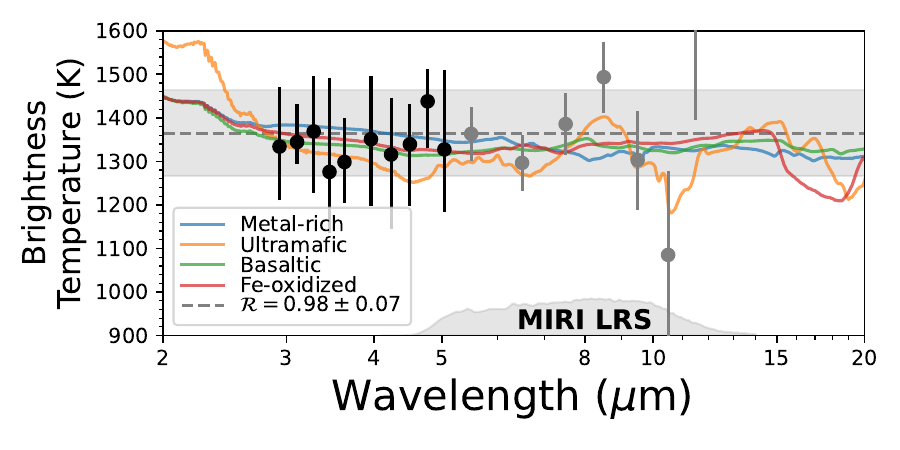}
    \caption{Model eclipse spectra of possible low-albedo surface types from \citet{hu12} in units of brightness temperature, ignoring the opposition surge effect. High brightness temperatures at low wavelengths are due to the contribution from reflected light. Black points represent our \eureka NIRSpec G395H eclipse spectrum, whereas grey points show simulated observations of 10 eclipses with MIRI LRS using \texttt{PandExo} \citep{pandexo} assuming an ultramafic surface.}
    \label{fig:surfaces}
\end{figure}

Based on the above analyses, TOI-1685~b likely does not have a significant atmosphere. For airless bodies, thermal emission can probe surface properties such as the regolith mineral composition \citep{hu12}. At a blackbody sub-stellar temperature of $\sim 1550$\,K, it is expected that the dayside of TOI-1685~b may be partially molten, although the majority is likely solid.  Molten and glass-quenched surfaces are expected to be dark and exhibit low albedo \citep{essack20}.  In addition, any regolith remaining on the planet would likely be extremely space-weathered by high-intensity stellar winds and micrometeorite impacts (e.g., \citealt{lyu24} and references therein), further darkening the surface albedo.

Despite its potential for identifying atmospheric spectral features, the NIRSpec wavelength range poses several challenges when attempting to distinguish between rocky planet surface types.  Unlike MIRI LRS, the wavelength range lacks the characteristic Si-O stretch absorption feature at 8--13\,µm \citep{hu12} and thus makes it difficult to distinguish between surface types in emission. The NIRSpec range also does not cover the 0.7--2.3\,µm range that contains possibly identifiable reflectance features for different surface types.  Therefore, if TOI-1685~b is a bare rock, our measurements can realistically only be used to constrain the overall surface albedo. While future eclipse observations may further constrain surface properties, as shown in Fig.~\ref{fig:surfaces}, a large number of visits with MIRI LRS will likely be needed to distinguish between surface types with statistical confidence.

\section{Conclusions} \label{sec:conclusions}

\begin{figure}
    \centering
    \includegraphics[width=1.0\linewidth]{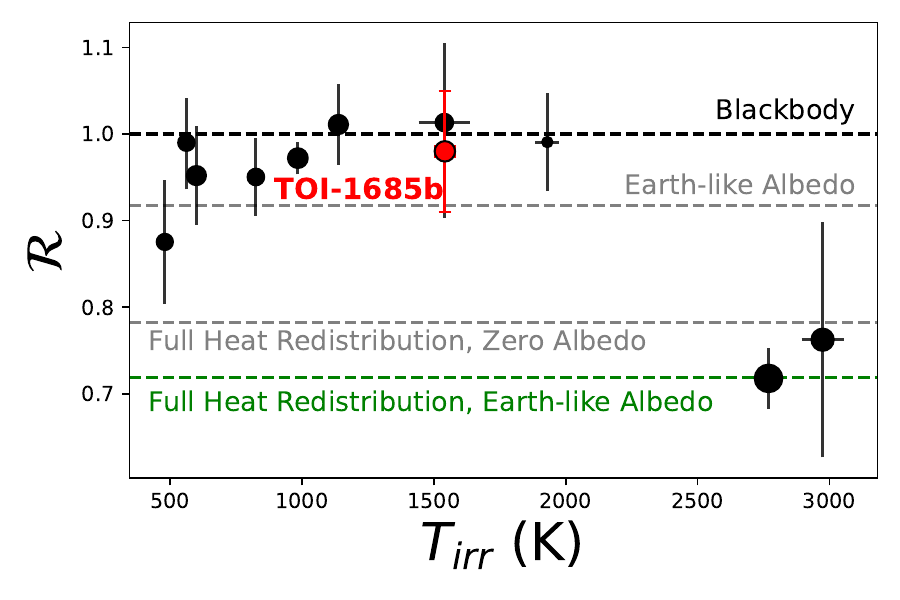}
    \caption{TOI-1685~b's measured NRS2 brightness temperature ratio $\mathcal{R}=0.98\pm0.07$ as a function of irradiation temperature ($T_{\rm irr}$, equivalent to the substellar temperature of a blackbody) in context of other rocky exoplanets observed in emission. Point size is proportional to planet size. TOI-1685~b follows the trend of blackbody-like emission for $\sim1000-2000\,\mathrm{K}$ planets, suggesting no significant atmospheres and dark surfaces. The most compelling evidence for thick atmospheres via heat redistribution has come from planets hot enough to vaporize rock --- namely the K-star planets 55~Cnc~e (\citealt{hu24}, but also non-detections/extreme variation in \citealt{Patel2024A&A...690A.159P} using NIRCam) and K2-141~b \citep[][using Spitzer]{zieba22}. Data are taken from, in order of increasing irradiation temperature: TRAPPIST-1~c \citep{zieba23}, TRAPPIST-1~b \citep{greene23}, LTT~1445~A~b \citep{Wachiraphan24}, GJ~1132~b \citep{xue24}, GJ~486~b \citep{mansfield24}, LHS~3844~b \citep{kreidberg19}, GJ~1252~b \citep{Crossfield22}, GJ~367~b \citep{zhang24}, 55~Cnc~e \citep{hu24}, and K2-141~b \citep{zieba22}.}
    \label{fig:R_Tirr}
\end{figure}

In this work, we analyze JWST NIRSpec/G395H phase curve observations of the hot super-Earth TOI-1685~b. The observations span 19\,hours continuously, covering two eclipses and one transit of the planet. The precise transit photometry allows us to refine the planet's radius, making TOI-1685~b consistent with having an Earth-like density rather than an iron-poor composition as previously reported. Although the individual precision on the transmission and emission spectrum is slightly better than the predictions from \texttt{PandExo}, the residuals of the white and spectroscopic light curves are dominated by a correlated noise component that appears to be independent of the data reduction pipeline used. Multiple attempts to mitigate this red noise at the data reduction level proved unsuccessful, while post-processing techniques such as PCA decorrelation remove information attributable to the orbital phase modulation that introduces degeneracies on the system properties. We use a prayer-bead analysis as a way to avoid underestimating the uncertainties in the fitted and derived parameters from the emission spectrum and white-light phase curve imposed by the correlated noise. The properties of this correlated noise appear to be detector-dependent, which may be an indication of a systematic origin rather than an astrophysical one. We encourage the community to investigate the origin of instrumental systematics of NIRSpec for long timeseries observations, especially when the signals of interest are of the order of 100\,ppm.

The analysis of the full phase curve, including the transmission and emission spectra of the planet, suggests that the properties of TOI-1685~b are consistent with a blackbody with no heat redistribution and very low albedo. In context with recent JWST observations, TOI-1685~b adds to the growing number of rocky planets around M stars with no or extremely thin atmospheres \citep{kreidberg19,Crossfield22,greene23,zieba23,zhang24,xue24,mansfield24,Wachiraphan24}. Figure~\ref{fig:R_Tirr} shows that our measured brightness temperature ratio $\mathcal{R}$ for TOI-1685~b is roughly consistent with the blackbody-like emission measured for other hot rocky planets around M~stars. Notably, TOI-1685~b is the first of these planets to use NIRSpec and one of the few with data during a full orbit.  Our high sensitivity to CO$_2$ spectral features and lack of detection further suggest that the planet is a bare, dark rock. MIRI LRS observations in emission could be used to probe possible surface composition and geology, but the number of visits would necessarily be high.

\bigskip

R.L., A.D.F., and S.E.M are supported by NASA through the NASA Hubble Fellowship grants HST-HF2-51559.001-A, HST-HF2-51530.001-A, and \#HST-HF2-51563.001-A, respectively, awarded by the Space Telescope Science Institute, which is operated by the Association of Universities for Research in Astronomy, Inc., for NASA, under contract NAS 5-26555.  
M.W.M. and M.Z. acknowledge support from the Heising-Simons Foundation through the 51 Pegasi b Fellowship Program. 
We acknowledge financial support from the Agencia Estatal de Investigaci\'on of the Ministerio de Ciencia e Innovaci\'on MCIN/AEI/10.13039/501100011033 and the ERDF “A way of making Europe” through project PID2021-125627OB-C32, and from the Centre of Excellence “Severo Ochoa” award to the Instituto de Astrofisica de Canarias. 

This work is based on observations made with the NASA/ESA/CSA James Webb Space Telescope. The data were obtained from the Mikulski Archive for Space Telescopes at the Space Telescope Science Institute, which is operated by the Association of Universities for Research in Astronomy, Inc., under NASA contract NAS 5-03127 for JWST. These observations are associated with program \#GO3263. The data is available at MAST: \dataset[10.17909/jhgr-hc58]{http://dx.doi.org/10.17909/jhgr-hc58}.


%

\vspace{5mm}
\facilities{JWST(NIRSpec)}


\software{
\texttt{Eureka!} \citep{eureka}, 
\texttt{Tiberius} \citep{Kirk2017,Kirk2021}, 
\texttt{ExoTIC-JEDI} \citep{alderson22,Alderson2023}, 
\texttt{ExoTIC-LD} \citep{exotic_ld},
\texttt{astropy} \citep{astropy22},
\texttt{dynesty} \citep{dynesty},
\texttt{emcee} \citep{mackey13}. }




\clearpage\newpage
\appendix

\section{Attempts to remove the correlated noise during data reduction} \label{app:preprocessing}

In this section, we summarize the tests carried out to reduce or eliminate the correlated noise seen in the phase curve observations. All these tests were carried out using \eureka as the data reduction pipeline.

\paragraph{Linearity checks}
We tested if detector nonlinearity effects were responsible for the correlated noise component. We try data reductions discarding the last group of NIRSpec integrations, analogously to previous studies using MIRI \citep{zhang24}. Our results with or without the last group are identical not only on the white light curve but also on the transmission and emission spectrum. The large number of groups per integration of our observations is very robust against nonlinearity effects or any other detector problems typically associated with bright targets and a low number of groups \citep{Birkmann2022A&A...661A..83B}.

\paragraph{Group level background subtraction}
The 1/$f$ noise in NIRSpec is one the most important sources of systematics for exoplanet studies \citep{Rustamkulov2022ApJ...928L...7R}. We experiment with and without background subtraction at the group level in Stage 1 of the \texttt{jwst} pipeline. The best results were obtained when correcting for the 1/$f$ noise at the group level, but it was not responsible for the hour-timescale correlated noise seen in the light curves. Regardless of the data reduction pipeline, which implements 1/$f$ noise corrections differently (see \citealt{Alderson2023} for \exotic, and \citealt{Kirk2021} for \tiberius), the results were identical.

\paragraph{Custom bias subtraction}
We also tested a custom bias subtraction within Stage 1 of \eureka, setting the \texttt{bias\_correction} flag to \texttt{None}, \texttt{mean}, and \texttt{smooth} (in this case \texttt{bias\_smooth\_length = 59}). None of the tests changed the properties of the correlated noise in the data.


\paragraph{Fine Guiding Sensor}
Finally, we explore the possibility that the correlated noise depends on the trace position movement on the detector or the pointing stability of the fine guiding sensor. We measure the trace position in $x$ and $y$ axes within Stage 3 of \eureka, but find no correlation with the red noise component. Including the $x$- and $y$-shifts of the trace in the phase curve model to decorrelate against them does not improve the fit. Similarly, the shifts in telescope pointing as measured by the fine guiding sensor do not correlate with the hour-timescale variability of the correlated noise in the data and were not used in the analysis.

\section{Additional Tables} \label{app:tables}

\begin{table}
    \centering
    \caption{NIRSpec/G395H NRS1 transmission spectrum.}
\begin{tabular}{l|c|c|c}
Wavelength (µm)   & \multicolumn{3}{c}{Transit depth (ppm)}  \\ 
                  & \texttt{Eureka!} & \texttt{Tiberius} & \texttt{ExoTIC-JEDI}            \\ \hline 
  2.8605 $\pm$ 0.0165 & $731.6_{-45.2}^{+45.1}$ & $741.6_{-50.8}^{+51.7}$ & $718.3_{-55.4}^{+52.0}$ \\
  2.8940 $\pm$ 0.0170 & $823.6_{-45.9}^{+44.8}$ & $840.2_{-50.3}^{+51.7}$ & $856.3_{-51.5}^{+49.2}$ \\
  2.9275 $\pm$ 0.0165 & $703.9_{-43.6}^{+41.0}$ & $748.9_{-47.7}^{+48.1}$ & $792.1_{-45.7}^{+41.8}$ \\
  2.9610 $\pm$ 0.0170 & $756.2_{-40.4}^{+39.5}$ & $702.5_{-45.1}^{+44.2}$ & $766.5_{-44.4}^{+41.4}$ \\
  2.9945 $\pm$ 0.0165 & $808.0_{-41.2}^{+40.0}$ & $811.2_{-44.7}^{+42.6}$ & $827.5_{-38.4}^{+36.7}$ \\
  3.0280 $\pm$ 0.0170 & $712.8_{-39.9}^{+38.6}$ & $771.0_{-45.9}^{+47.3}$ & $723.9_{-50.6}^{+47.5}$ \\
  3.0615 $\pm$ 0.0165 & $864.7_{-39.2}^{+40.3}$ & $889.7_{-42.9}^{+43.6}$ & $898.9_{-32.7}^{+31.0}$ \\
  3.0950 $\pm$ 0.0170 & $785.7_{-40.0}^{+39.2}$ & $780.3_{-44.3}^{+42.9}$ & $741.1_{-40.3}^{+37.8}$ \\
  3.1285 $\pm$ 0.0165 & $774.9_{-38.7}^{+39.0}$ & $777.9_{-43.4}^{+43.4}$ & $764.9_{-50.5}^{+46.8}$ \\
  3.1620 $\pm$ 0.0170 & $804.0_{-39.1}^{+38.5}$ & $766.5_{-44.8}^{+45.2}$ & $757.7_{-45.2}^{+41.1}$ \\
  3.1955 $\pm$ 0.0165 & $761.8_{-38.0}^{+36.7}$ & $753.3_{-41.7}^{+41.8}$ & $758.1_{-35.7}^{+33.6}$ \\
  3.2290 $\pm$ 0.0170 & $862.8_{-39.3}^{+38.1}$ & $853.1_{-43.7}^{+45.1}$ & $892.8_{-42.2}^{+38.9}$ \\
  3.2625 $\pm$ 0.0165 & $788.7_{-39.3}^{+40.2}$ & $744.1_{-43.6}^{+42.6}$ & $798.6_{-35.7}^{+33.5}$ \\
  3.2960 $\pm$ 0.0170 & $759.1_{-38.9}^{+37.8}$ & $766.8_{-43.3}^{+43.5}$ & $774.2_{-40.8}^{+38.2}$ \\
  3.3295 $\pm$ 0.0165 & $841.9_{-38.5}^{+38.8}$ & $826.3_{-43.3}^{+45.5}$ & $867.8_{-45.1}^{+42.6}$ \\
  3.3630 $\pm$ 0.0170 & $827.5_{-38.7}^{+38.2}$ & $825.4_{-42.9}^{+43.1}$ & $837.0_{-34.1}^{+32.1}$ \\
  3.3965 $\pm$ 0.0165 & $854.8_{-40.2}^{+40.3}$ & $821.6_{-44.6}^{+44.3}$ & $836.8_{-41.0}^{+37.8}$ \\
  3.4300 $\pm$ 0.0170 & $804.5_{-38.5}^{+38.0}$ & $816.3_{-42.9}^{+41.3}$ & $822.7_{-35.9}^{+34.3}$ \\
  3.4640 $\pm$ 0.0170 & $734.7_{-38.0}^{+37.9}$ & $720.0_{-42.2}^{+43.4}$ & $773.5_{-38.0}^{+34.9}$ \\
  3.4975 $\pm$ 0.0165 & $817.7_{-40.0}^{+39.9}$ & $825.9_{-44.7}^{+42.6}$ & $795.9_{-40.1}^{+37.9}$ \\
  3.5310 $\pm$ 0.0170 & $760.6_{-40.5}^{+39.5}$ & $735.1_{-44.1}^{+44.6}$ & $786.7_{-37.9}^{+35.7}$ \\
  3.5645 $\pm$ 0.0165 & $744.3_{-38.2}^{+40.1}$ & $748.4_{-46.3}^{+46.7}$ & $728.9_{-44.8}^{+42.7}$ \\
  3.5980 $\pm$ 0.0170 & $774.7_{-39.0}^{+38.7}$ & $762.7_{-43.2}^{+45.6}$ & $743.2_{-64.7}^{+60.2}$ \\
  3.6315 $\pm$ 0.0165 & $810.1_{-39.2}^{+39.9}$ & $820.1_{-45.1}^{+45.0}$ & $827.1_{-41.4}^{+39.3}$ \\
  3.6650 $\pm$ 0.0170 & $775.7_{-40.6}^{+39.6}$ & $746.3_{-44.7}^{+48.0}$ & $769.4_{-42.0}^{+39.9}$ \\
  3.6985 $\pm$ 0.0165 & $811.8_{-42.1}^{+41.5}$ & $802.0_{-43.9}^{+45.1}$ & $775.1_{-42.2}^{+40.0}$ \\ \hline
\end{tabular}
\label{tab:transmission_nrs1}
\end{table}
  
\begin{table}
    \centering
    \caption{NIRSpec/G395H NRS2 transmission spectrum.}
\begin{tabular}{l|c|c|c}
Wavelength (µm)   & \multicolumn{3}{c}{Transit depth (ppm)}  \\ 
                  & \texttt{Eureka!} & \texttt{Tiberius} & \texttt{ExoTIC-JEDI}            \\ \hline 
  3.8400 $\pm$ 0.0170 & $745.5_{-44.4}^{+44.7}$ & $780.8_{-52.2}^{+52.3}$ & $713.7_{-52.3}^{+48.6}$ \\
  3.8735 $\pm$ 0.0165 & $743.9_{-45.4}^{+46.1}$ & $728.7_{-53.8}^{+54.0}$ & $740.0_{-58.5}^{+53.0}$ \\
  3.9070 $\pm$ 0.0170 & $753.5_{-44.8}^{+43.9}$ & $781.9_{-52.9}^{+56.4}$ & $725.9_{-60.6}^{+54.5}$ \\
  3.9410 $\pm$ 0.0170 & $754.7_{-47.5}^{+47.1}$ & $772.0_{-54.1}^{+54.6}$ & $682.8_{-62.3}^{+57.6}$ \\
  3.9750 $\pm$ 0.0170 & $771.6_{-47.8}^{+47.3}$ & $688.6_{-63.4}^{+65.4}$ & $783.2_{-50.5}^{+46.8}$ \\
  4.0085 $\pm$ 0.0165 & $894.8_{-48.3}^{+49.4}$ & $954.3_{-70.1}^{+68.9}$ & $914.0_{-55.5}^{+53.6}$ \\
  4.0420 $\pm$ 0.0170 & $688.9_{-50.1}^{+50.1}$ & $670.9_{-61.4}^{+61.5}$ & $658.6_{-87.7}^{+79.2}$ \\
  4.0760 $\pm$ 0.0170 & $801.4_{-49.9}^{+51.4}$ & $809.1_{-61.1}^{+63.2}$ & $817.3_{-65.0}^{+61.0}$ \\
  4.1095 $\pm$ 0.0165 & $725.7_{-49.5}^{+51.5}$ & $740.4_{-59.7}^{+59.9}$ & $777.7_{-68.5}^{+64.5}$ \\
  4.1430 $\pm$ 0.0170 & $718.2_{-50.7}^{+51.1}$ & $717.6_{-55.2}^{+55.4}$ & $659.4_{-76.3}^{+69.5}$ \\
  4.1770 $\pm$ 0.0170 & $715.1_{-52.9}^{+53.1}$ & $749.1_{-63.2}^{+64.3}$ & $754.3_{-73.6}^{+67.2}$ \\
  4.2110 $\pm$ 0.0170 & $706.6_{-51.9}^{+53.7}$ & $727.1_{-58.0}^{+58.4}$ & $693.3_{-87.3}^{+80.5}$ \\
  4.2445 $\pm$ 0.0165 & $812.8_{-56.7}^{+59.0}$ & $781.7_{-58.6}^{+60.4}$ & $821.0_{-90.4}^{+84.6}$ \\
  4.2780 $\pm$ 0.0170 & $867.5_{-55.8}^{+53.1}$ & $825.5_{-60.3}^{+58.8}$ & $724.1_{-95.7}^{+87.8}$ \\
  4.3120 $\pm$ 0.0170 & $729.4_{-58.5}^{+59.1}$ & $723.1_{-63.7}^{+61.4}$ & $739.9_{-97.7}^{+87.2}$ \\
  4.3460 $\pm$ 0.0170 & $771.7_{-59.0}^{+59.7}$ & $732.0_{-64.9}^{+63.4}$ & $693.3_{-106}^{+94.0}$ \\
  4.3795 $\pm$ 0.0165 & $776.3_{-65.4}^{+65.1}$ & $738.6_{-66.6}^{+68.1}$ & $764.5_{-109}^{+95.6}$ \\
  4.4130 $\pm$ 0.0170 & $750.1_{-62.0}^{+64.7}$ & $803.3_{-67.9}^{+68.5}$ & $660.5_{-133}^{+120}$  \\
  4.4470 $\pm$ 0.0170 & $721.5_{-65.7}^{+67.6}$ & $804.7_{-70.4}^{+71.4}$ & $675.6_{-129}^{+113}$  \\
  4.4810 $\pm$ 0.0170 & $662.8_{-68.9}^{+68.4}$ & $666.0_{-73.2}^{+71.6}$ & $671.9_{-147}^{+130}$  \\
  4.5145 $\pm$ 0.0165 & $866.5_{-68.6}^{+70.3}$ & $831.5_{-71.7}^{+75.2}$ & $828.5_{-112}^{+103}$  \\
  4.5480 $\pm$ 0.0170 & $848.2_{-68.7}^{+70.5}$ & $844.4_{-74.8}^{+78.6}$ & $987.5_{-105}^{+99.5}$  \\
  4.5820 $\pm$ 0.0170 & $786.6_{-72.8}^{+71.6}$ & $704.3_{-75.0}^{+77.0}$ & $795.2_{-131}^{+115}$  \\
  4.6155 $\pm$ 0.0165 & $793.7_{-77.9}^{+75.3}$ & $806.8_{-88.0}^{+92.4}$ & $871.9_{-129}^{+114}$  \\
  4.6490 $\pm$ 0.0170 & $833.3_{-71.9}^{+74.4}$ & $818.2_{-79.6}^{+80.1}$ & $733.5_{-142}^{+129}$  \\
  4.6830 $\pm$ 0.0170 & $778.2_{-74.3}^{+75.8}$ & $751.8_{-82.4}^{+79.4}$ & $923.3_{-119}^{+107}$  \\
  4.7170 $\pm$ 0.0170 & $895.4_{-73.6}^{+75.9}$ & $959.7_{-85.6}^{+84.9}$ & $786.0_{-138}^{+120}$  \\
  4.7505 $\pm$ 0.0165 & $740.1_{-79.5}^{+79.7}$ & $737.2_{-89.6}^{+89.2}$ & $700.9_{-160}^{+142}$  \\
  4.7840 $\pm$ 0.0170 & $718.1_{-79.6}^{+77.1}$ & $702.4_{-90.3}^{+89.6}$ & $748.1_{-151}^{+134}$  \\
  4.8180 $\pm$ 0.0170 & $772.1_{-80.9}^{+82.1}$ & $705.2_{-89.0}^{+91.3}$ & $800.8_{-139}^{+126}$  \\
  4.8520 $\pm$ 0.0170 & $666.7_{-82.3}^{+81.7}$ & $655.0_{-88.6}^{+91.6}$ & $683.8_{-168}^{+148}$  \\
  4.8855 $\pm$ 0.0165 & $785.7_{-84.8}^{+86.8}$ & $785.3_{-91.8}^{+94.5}$ & $771.3_{-153}^{+138}$  \\
  4.9190 $\pm$ 0.0170 & $805.5_{-83.1}^{+83.1}$ & $719.3_{-94.2}^{+94.0}$ & $805.6_{-144}^{+131}$  \\
  4.9530 $\pm$ 0.0170 & $751.9_{-87.3}^{+88.5}$ & $757.4_{-102 }^{+103 }$ & $716.8_{-168}^{+148}$  \\
  4.9865 $\pm$ 0.0165 & $853.8_{-86.3}^{+87.3}$ & $812.0_{-100 }^{+105 }$ & $913.7_{-133}^{+118}$  \\
  5.0200 $\pm$ 0.0170 & $764.4_{-87.5}^{+88.6}$ & $689.7_{-125 }^{+126 }$ & $614.6_{-201}^{+171}$  \\
  5.0540 $\pm$ 0.0170 & $597.2_{-92.4}^{+86.8}$ & $430.1_{-123 }^{+122 }$ & $535.7_{-229}^{+195}$  \\
  5.0880 $\pm$ 0.0170 & $916.5_{-93.5}^{+90.9}$ & $853.7_{-120 }^{+125 }$ & $1008._{-121}^{+108}$  \\
  5.1215 $\pm$ 0.0165 & $738.5_{-96.6}^{+94.6}$ & $642.2_{-125 }^{+119 }$ & $690.9_{-178}^{+157}$  \\
  5.1550 $\pm$ 0.0170 & $829.3_{-102.}^{+100.}$ & $757.6_{-125 }^{+127 }$ & $616.9_{-214}^{+188}$  \\ \hline
\end{tabular}
\label{tab:transmission_nrs2}
\end{table}

\begin{table*}[h!]
\begin{rotatetable*}
    \centering
    \caption{NIRSpec/G395H emission spectrum for each reduction and dataset shown in Fig.~\ref{fig:emission_all}.}
\begin{tabular}{c|rrrr|rr|rrr}
Wavelength (µm)   & \multicolumn{9}{c}{Eclipse depth (ppm)}  \\ 
                  & \multicolumn{4}{c}{\texttt{Eureka!}} & \multicolumn{2}{c}{\texttt{Tiberius}} & \multicolumn{3}{c}{\texttt{ExoTIC-JEDI}}            \\ 
                  & First & Second & Combined & Full     & Combined & Full                       & First & Second & Combined            \\ \hline 
2.9310 $\pm$ 0.087 & $ 75.1_{-21.8}^{+21.4}$ & $ 69.0_{-29.7}^{+28.9}$ & $ 58.7_{-16.6}^{+17.1}$ & $ 91.5_{-10.0}^{+10.0}$ & $ 54.7_{-16.7}^{+16.7}$ & $ 93.5_{-10.5}^{+10.7}$ & $ 86.7\pm20.3$ & $ 57.1\pm16.2$ & $ 77.7\pm17.9$ \\
3.1050 $\pm$ 0.087 & $ 82.1_{-20.0}^{+19.3}$ & $ 94.6_{-26.6}^{+26.9}$ & $ 70.7_{-14.8}^{+14.6}$ & $ 96.7_{-10.0}^{+10.4}$ & $ 72.0_{-15.0}^{+14.9}$ & $101.0_{-10.0}^{+10.1}$ & $ 96.2\pm21.6$ & $ 86.1\pm20.0$ & $ 78.5\pm17.7$ \\ 
3.2795 $\pm$ 0.087 & $102.4_{-20.2}^{+19.2}$ & $ 88.8_{-26.8}^{+27.8}$ & $ 76.2_{-15.1}^{+14.9}$ & $114.9_{- 9.9}^{+ 9.6}$ & $ 70.9_{-16.2}^{+16.1}$ & $111.4_{-10.6}^{+10.5}$ & $ 86.0\pm20.0$ & $ 66.3\pm16.9$ & $ 75.4\pm17.3$ \\ 
3.4540 $\pm$ 0.087 & $108.2_{-18.3}^{+18.9}$ & $ 63.9_{-26.7}^{+26.6}$ & $ 63.5_{-14.7}^{+14.5}$ & $104.0_{- 9.5}^{+ 9.7}$ & $ 64.4_{-15.7}^{+15.5}$ & $ 97.9_{-10.0}^{+10.4}$ & $ 99.8\pm22.1$ & $ 43.5\pm13.9$ & $ 73.1\pm16.9$ \\ 
3.6280 $\pm$ 0.087 & $ 77.3_{-21.1}^{+20.3}$ & $ 78.8_{-26.1}^{+27.5}$ & $ 64.8_{-15.3}^{+15.2}$ & $110.6_{-10.2}^{+10.4}$ & $ 66.3_{-16.0}^{+15.7}$ & $113.5_{-10.8}^{+10.8}$ & $ 94.1\pm21.4$ & $ 61.6\pm16.4$ & $ 78.1\pm17.8$ \\ \hline
3.9580 $\pm$ 0.135 & $ 65.6_{-17.7}^{+17.4}$ & $ 90.4_{-19.2}^{+20.8}$ & $ 78.7_{-13.6}^{+13.5}$ & $ 96.0_{-10.4}^{+10.5}$ & $ 98.0_{-16.3}^{+16.9}$ & $109.3_{-12.9}^{+13.5}$ & $ 90.7\pm21.0$ & $ 57.2\pm16.3$ & $ 81.1\pm18.6$ \\
4.2280 $\pm$ 0.135 & $ 59.8_{-20.5}^{+19.3}$ & $ 96.3_{-22.5}^{+21.2}$ & $ 79.2_{-15.1}^{+15.9}$ & $113.1_{-11.8}^{+12.0}$ & $ 86.8_{-16.6}^{+17.1}$ & $123.6_{-13.0}^{+13.7}$ & $ 77.9\pm18.6$ & $ 88.8\pm20.6$ & $ 83.6\pm18.8$ \\ 
4.4975 $\pm$ 0.135 & $ 86.5_{-23.6}^{+24.0}$ & $114.3_{-28.2}^{+28.2}$ & $100.3_{-19.1}^{+19.4}$ & $131.0_{-14.8}^{+14.7}$ & $106.4_{-22.0}^{+20.6}$ & $132.3_{-16.1}^{+16.9}$ & $100.1\pm23.0$ & $120.5\pm27.0$ & $109.5\pm24.0$  \\ 
4.7700 $\pm$ 0.135 & $139.1_{-25.2}^{+25.1}$ & $148.0_{-28.7}^{+28.9}$ & $138.2_{-20.9}^{+21.2}$ & $178.3_{-16.1}^{+16.3}$ & $137.6_{-22.4}^{+24.0}$ & $183.1_{-17.8}^{+18.1}$ & $124.4\pm27.7$ & $112.2\pm25.8$ & $118.7\pm26.0$ \\ 
5.0370 $\pm$ 0.135 & $ 90.7_{-28.3}^{+29.5}$ & $142.9_{-31.1}^{+30.9}$ & $122.4_{-24.3}^{+23.8}$ & $151.1_{-18.5}^{+19.1}$ & $121.1_{-29.2}^{+29.0}$ & $143.0_{-24.6}^{+24.9}$ & $ 85.3\pm22.0$ & $152.2\pm33.5$ & $118.7\pm26.5$ \\ \hline
\end{tabular}
\label{tab:emission}
\end{rotatetable*}
\end{table*}

\section{Additional Figures} \label{app:figures}

\begin{figure}
    \centering
    \includegraphics[width=0.8\linewidth]{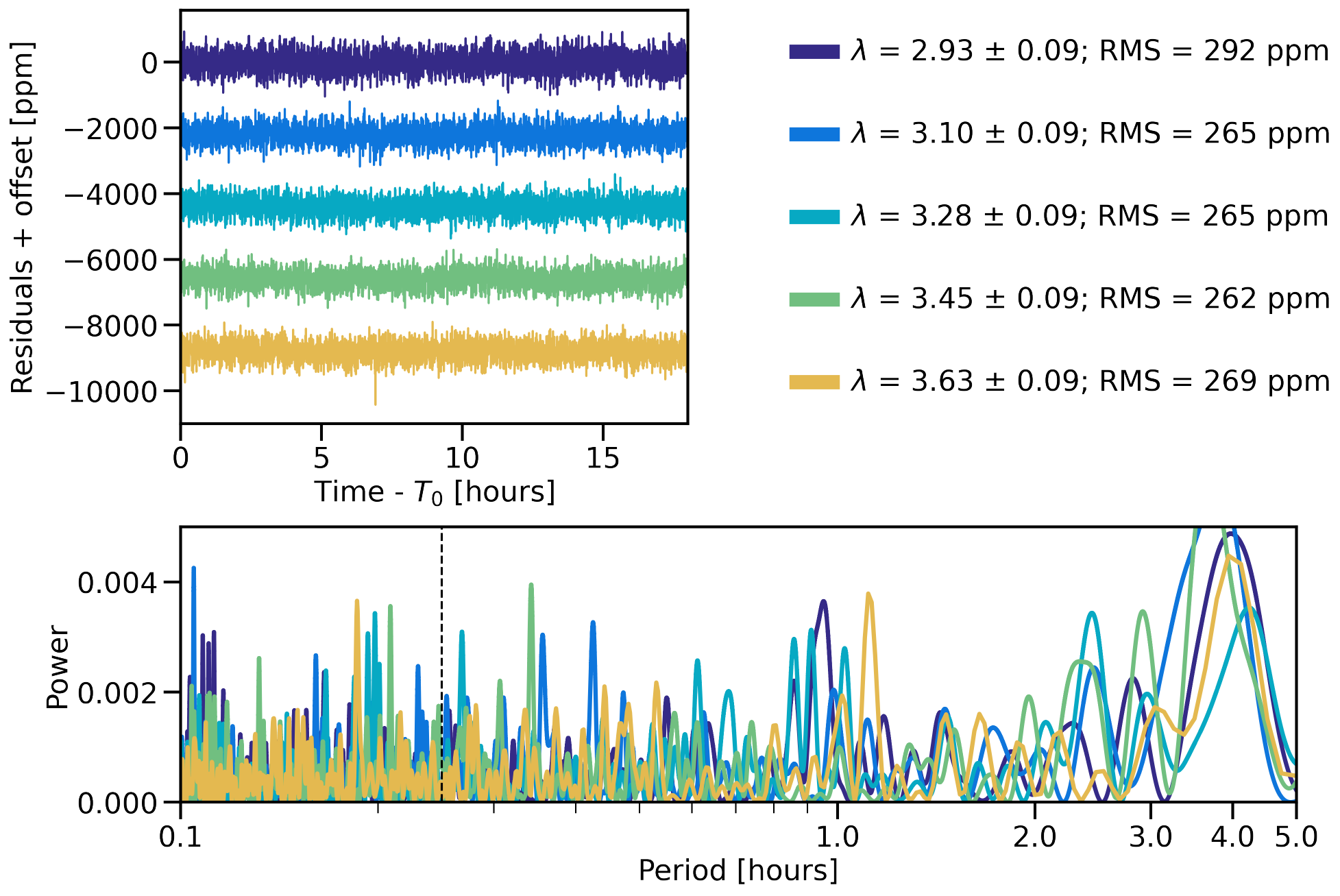}
    \caption{Results of running the residuals of the white-light and spectroscopic light curves through a Lomb-Scargle periodogram. (a) A gallery of the residuals from the \texttt{Eureka!} pipeline for each spectroscopic channel. (b) The periodogram results. The color of the lines between panels a and b are for the same white-light or spectroscopic channel. The residuals show uncorrelated noise at periods of $< 0.5$~hours.}
    \label{fig:nrs1_ls}
\end{figure}

\begin{figure}
    \centering
    \includegraphics[width=0.8\linewidth]{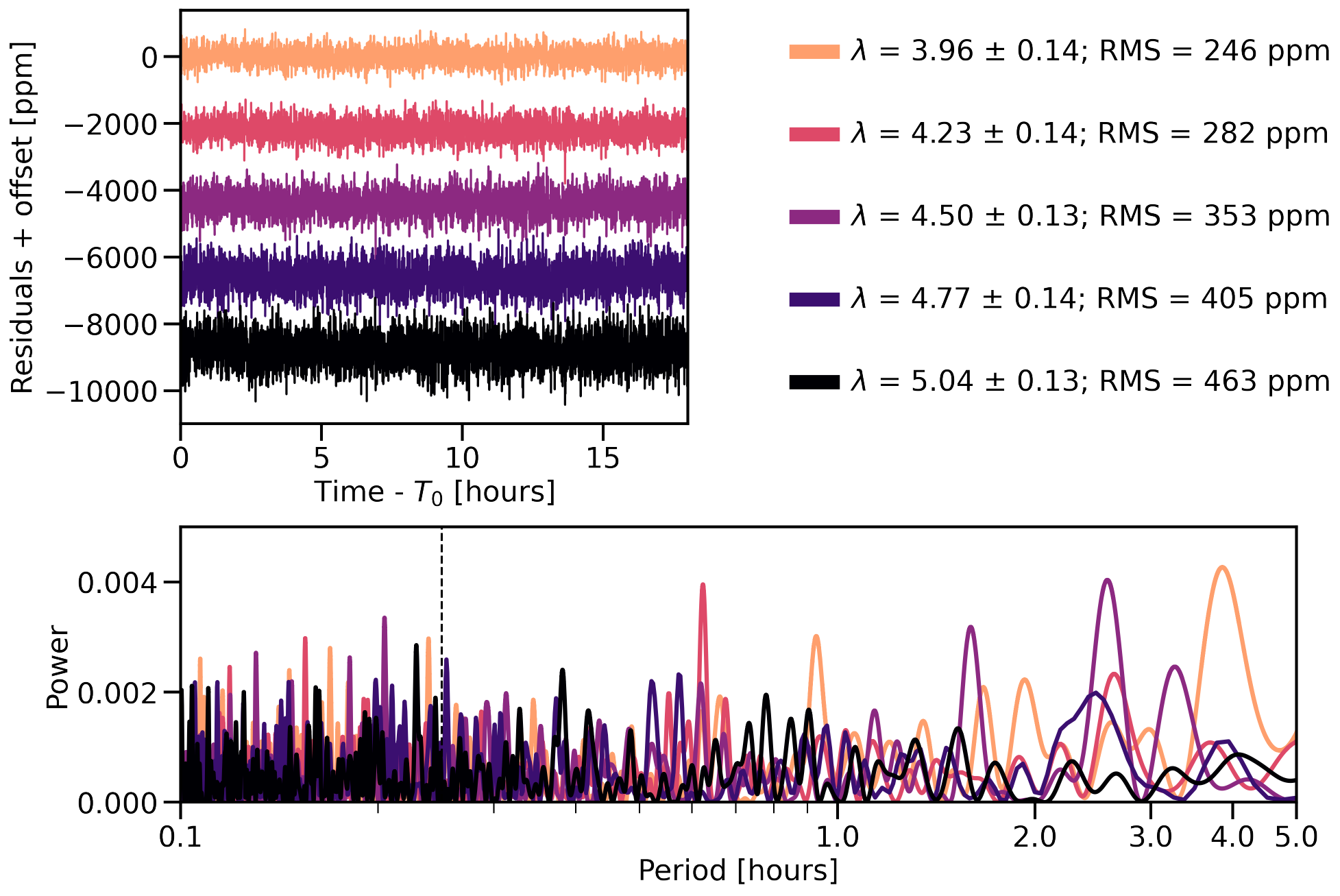}
    \caption{Same as Figure~\ref{fig:nrs1_ls}, but for NRS2.}
    \label{fig:nrs2_ls}
\end{figure}

\begin{figure*}
    \centering
    \includegraphics[width=1.0\linewidth]{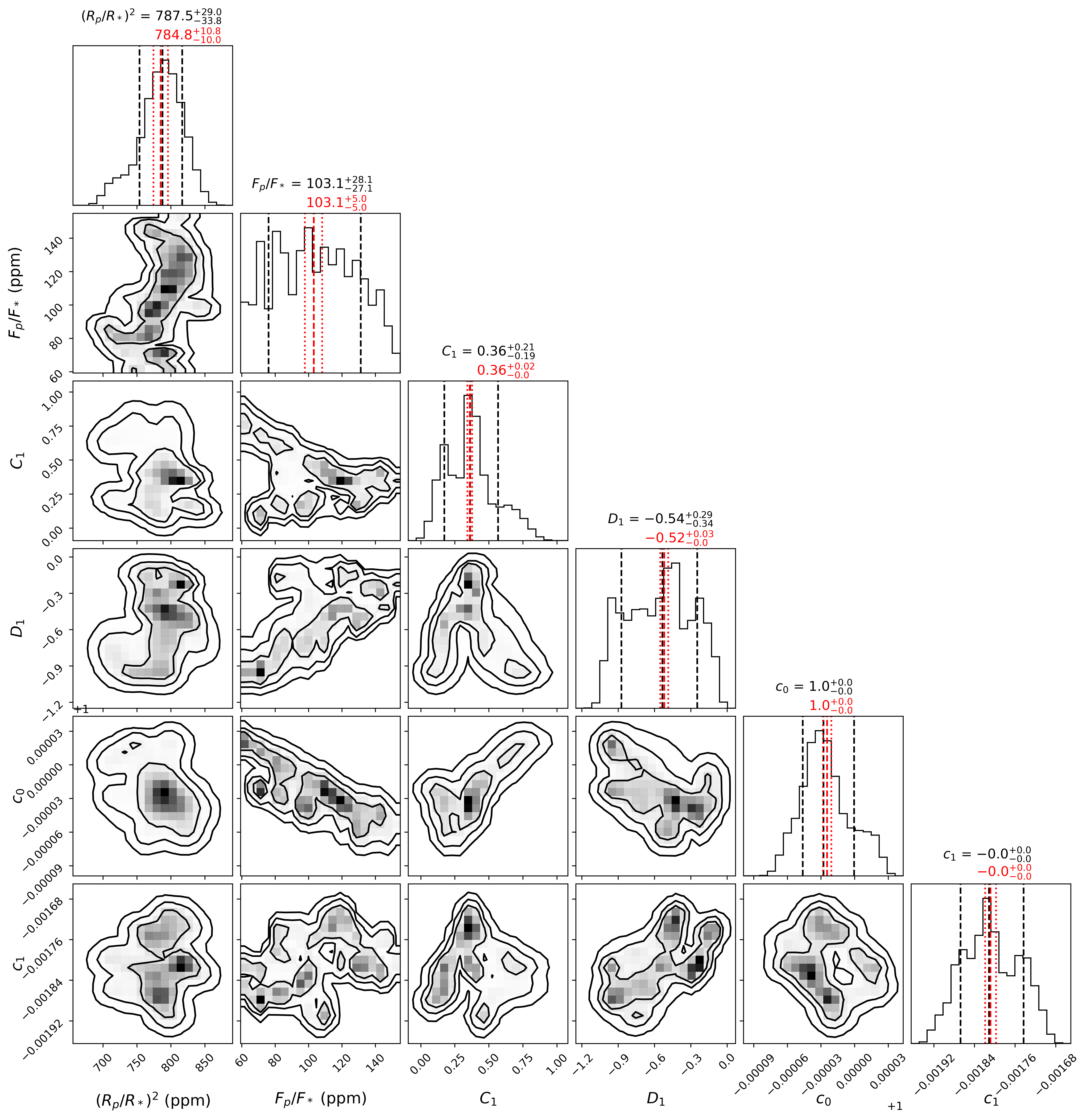}
    \caption{Posterior distributions of fitted parameters from the NRS1 white light curve analysis. As discussed in \S \ref{subsubsec:prayerbead}, we performed the prayer-bead method on the \texttt{Eureka!} reduction with MCMC, resulting in  4370 values of median and $\pm$1$\sigma$. We then re-created 4370 Gaussian distributions using these medians and $\pm$1$\sigma$, each with 1000 points, and plotted the 1,2,3$\sigma$ contours of the distributions on this figure. The median and $\pm1\sigma$ of the distributions are shown with black dashed and dotted lines, while the original results from \texttt{Eureka!} analysis are shown with red lines. Clearly, the prayer-bead analysis provides us with larger errorbars, thus avoiding underestimations of the true uncertainties. }
    \label{fig:corner_prayerbead_nrs1}
\end{figure*}

\begin{figure*}
    \centering
    \includegraphics[width=1.0\linewidth]{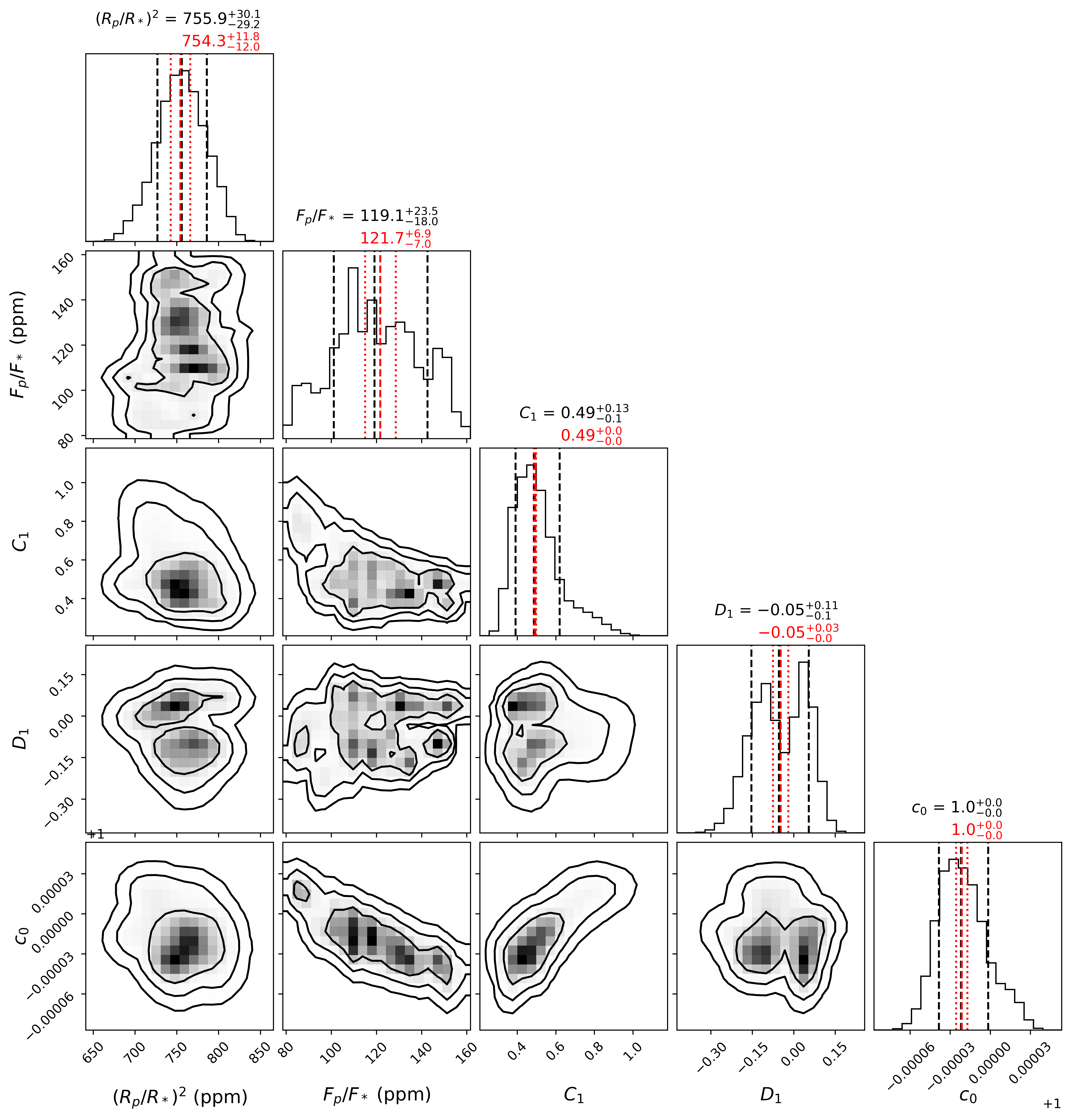}
    \caption{Posterior distributions of fitted parameters from the NRS2 white light curve analysis. See the caption of Fig.\ref{fig:corner_prayerbead_nrs1} for more information.}
    \label{fig:corner_prayerbead_nrs2}
\end{figure*}



\clearpage\newpage
\bibliography{main}
\bibliographystyle{aasjournal}



\end{document}